\documentclass[final,onefignum,onetabnum]{siamonline190516}
\usepackage{amssymb}
\usepackage{subcaption}
\usepackage{psfrag}
\usepackage{physics}

\usepackage{tabularx}
\newcommand{\Poincare}{Poincar\'e}
\newcommand{\Rey}{\mathrm{Re}}

\usepackage{lipsum}
\usepackage{amsfonts}
\usepackage{graphicx}
\usepackage{epstopdf}
\usepackage{algorithmic}
\ifpdf
  \DeclareGraphicsExtensions{.eps,.pdf,.png,.jpg}
\else
  \DeclareGraphicsExtensions{.eps}
\fi

\usepackage{enumitem}
\setlist[enumerate]{leftmargin=.5in}
\setlist[itemize]{leftmargin=.5in}

\newsiamremark{remark}{Remark}
\newsiamremark{hypothesis}{Hypothesis}
\crefname{hypothesis}{Hypothesis}{Hypotheses}
\newsiamthm{claim}{Claim}

\headers{Dynamics of small particle inertial migration in curved square ducts}{Kyung Ha, Brendan Harding, Yvonne M. Stokes,  Andrea L. Bertozzi} % BH: suggestion for shorter running title?

\title{Dynamics of small particle inertial migration in curved square ducts%
\thanks{Submitted to the editors DATE.
\funding{This research is supported under Australian Research Council’s Discovery Projects funding scheme (DP160102021 and DP200100834), an Australian Research Council Future Fellowship (FT160100108) to YMS, and the Simons Foundation Math + X Investigator Award 510776.}
}}

\author{Kyung Ha\thanks{Department of Mathematics, University
of California, Los Angeles, CA USA (\email{kyungha@g.ucla.edu},\email{bertozzi@math.ucla.edu}).}
\and Brendan Harding\thanks{School of Mathematics and Statistics, Victoria University of Wellington, New Zealand (\email{brendan.harding@vuw.ac.nz}).}
\and Andrea L. Bertozzi$^\dagger$\thanks{Department of Mechanical and Aerospace Engineering, University
of California, Los Angeles, CA USA.}
\and Yvonne M. Stokes\thanks{School of Mathematical Sciences, University
of Adelaide, South Australia, Australia (\email{yvonne.stokes@adelaide.edu.au}).}
}

\usepackage{amsopn}

\ifpdf
\hypersetup{
  pdftitle={The dynamics of small particles undergoing inertial migration in curved square ducts}, % BH: suggestion/placeholder
  pdfauthor={Kyung Ha, Brendan Harding, Yvonne M. Stokes,  Andrea L. Bertozzi}
}
\fi

\newcommand{\ys}[1]{#1}
\newcommand{\bh}[1]{#1}
\newcommand{\kh}[1]{#1}

\newcommand{\red}[1]{#1}
\begin{document}

\maketitle
% REQUIRED

\begin{abstract}
Microchannels are well-known in microfluidic applications for the control and separation of microdroplets and cells.  
Often the objects in the flow experience inertial effects, resulting in dynamics that is a departure from the underlying channel flow dynamics.  
This paper considers small neutrally buoyant spherical particles suspended in flow through a curved duct having a square cross-section.
The particle experiences a combination of inertial lift force induced by the disturbance from the primary flow along the duct, and drag from the secondary vortices in the cross-section,  which drive migration of the particle within the cross-section.
We construct a simplified model that preserves the core topology of the force field yet
depends on a single parameter $\kappa$, quantifying the relative strength of the two forces. We show that $\kappa$ is a bifurcation parameter for the dynamical system that describes motion of the particle in the cross section of the duct. At large values of $\kappa$ there exists an attracting limit cycle, in each of the upper and lower halves of the duct. At small $\kappa$ we find that particles migrate to one of four stable foci. Between these extremes, there is an intermediate-range of $\kappa$ for which all particles migrate to a single stable focus. Noting that the positions of the limit cycles and foci vary with the value of $\kappa$, this behavior indicates that, for a suitable particle mixture, duct bend radius might be chosen to segregate particles by size.
We evaluate the time and axial distance required to focus particles near the unique stable node, which determines the duct length required for particle segregation.
\end{abstract}

% REQUIRED
\begin{keywords}
inertial migration; limit cycle; microfluidics
\end{keywords}

% REQUIRED
\begin{AMS}
  	37C27,70K05,	70K42,	70K70,76D05
\end{AMS}

\section{Introduction}

% PREVIOUS WORK AND APPLICATIONS
%%% BH: The following paragraph is adapted from my DECRA application, may as well get some use from it!
Inertial lift is a phenomenon, first reported by Segre and Silberberg~\cite{SegreSilberberg1961}, that causes particles and cells suspended in flow through microscale devices to deviate from fluid streamlines.
The applications of this effect are revolutionizing diagnostic medical technologies, the separation and identification of circulating tumor cells (CTCs) being one of many examples~\cite{WarkianiEtal2014}.
Other general uses include flow cytometry, rare cell isolation, cell cycle synchronization, platelet and bacteria separation, plasma extraction, particle classification, and fluid mixing~\cite{MartelToner2014}.
While the effect of inertial lift on particle migration has been extensively studied for uni-directional flows~\cite{Asmolov1999,HoLeal1974,Hood2015JFM,MatasMG2004,SchonbergHinch1989}, many of the devices being used in cutting edge microfluidics have a complex design through which there is a full three-dimensional flow. 
Some of the more recent advances in this field, experimental and otherwise, are described in several review articles~\cite{DiCarlo2009,GeislingerFranke2014,ShiRzehak2020}.

% SUMMARY OF PREVIOUS WORK
Inertial migration of a neutrally buoyant spherical particle suspended in flow through a straight duct with square cross-section was considered by Hood et. al~\cite{Hood2015JFM}.
Analysis of the fully enclosed flow is challenging, and to render the problem tractable a combination of perturbation theory and numerical computation was applied.
Motivated by their approach, Harding et al.~\cite{Harding2019JFM} extended this work to consider the inertial migration of a neutrally buoyant spherical particle suspended in flow through curved ducts with square, rectangular and trapezoidal cross-sections; 
it was found that rectangular and trapezoidal cross-sections had a better ability to separate particles depending on their size and these cross-sections became the primary focus of the results presented.
In particular, it was demonstrated that the lateral focusing location within curved ducts with a rectangular cross-section could be approximately characterized by \ys{a} dimensionless parameter $\kappa$ which \ys{approximately quantifies} the relative \ys{strength} of the secondary flow drag to the inertial lift force, these being the primary drivers of particle migration.
Although the migration of only a single particle was considered it is reasonable to expect \ys{good predictions} of the behavior of sufficiently dilute suspensions in which particle-particle interaction is minimal.
\ys{In this paper we focus on curved ducts with a square cross-section which, as seen in \cite{Harding2019JFM}, exhibit} a variety of interesting migration dynamics that warrant further investigation.
This is one aim of the present paper.

%% TODO: add a paragraph or two that talks a little about previous efforts to model the inertial lift force, e.g. 'curl models' and such, and how they don't work particularly well, thus motivating our approach for constructing the ZeLF model (which preserves the `key topological properties').
A second aim is to construct a model of the migration forces on a particle \ys{in a curved duct} which is simple to evaluate whilst still capturing the topological structure which is essential for accurate prediction of migration dynamics.
While simulation data from \cite{Harding2019JFM} can be interpolated directly and applied to the integration of particle trajectories, a simpler closed-form model facilitates rapid prototyping. 
Existing models in the literature often focus on modeling the inertial lift force as a sum of \ys{wall-induced,
%shear-gradient-induced,
slip-shear} and shear-gradient-induced components.
Such models are generally one-dimensional in nature owing to the historical development of this decomposition via a study of particle migration in one-dimensional flows between two plane parallel walls.
Rasooli and \c{C}etin~\cite{RasooliCetin2018} \ys{remark that the application of such models} ``for the prediction of equilibria for particles in 3D Poiseuille flow in square and rectangular channels is quite questionable''.
They \ys{instead} use Hood et al.'s approximation of the inertial lift force in straight rectangular ducts~\cite{Hood2015JFM} for their own particle tracking model applied to flow through curved rectangular ducts.
The idea of combining the inertial lift force from a straight duct with drag forces induced from curved duct flow \ys{as a simple but useful way to describe behavior in curved ducts} was also explored by Harding~\cite{Harding2018AFMC}.

%% BH: probably not needed
% As another example, Liu et al. studied migration in rectangular channels with large aspect ratio using an inertial lift model acting only in the vertical direction~\cite{LiuEtal2016}.
% Despite missing horizontal components of the inertial lift force it showed some qualitative agreement when applied to two complex channel geometries which each featured a non-constant secondary fluid flow having a significant impact on the migration dynamics (which may have lessened the need for the horizontal component).
% It is clear that such models do not capture the correct topological structure of the inertial lift force which is essential to accurately reproducing migration dynamics more generally.

Our particle tracking model for neutrally buoyant spherical particles suspended in flow through curved square ducts is similar to that of Rasooli and \c{C}etin~\cite{RasooliCetin2018} but differs in a few key ways: a) the axial particle velocity is taken to be in constant equilibrium with the surrounding fluid, b) the forces within a cross-section are decomposed solely into a secondary flow drag and inertial lift force, and c) both the secondary flow drag and inertial lift force are modeled via relatively simple formulae.
The assumption in a), which includes neglect of the added mass force, is reasonable because equilibrium in the main direction of flow is reached quickly compared to the time scale of particle migration. 
Moreover, as the particle migrates in the cross-sectional plane the change in axial velocity is sufficiently smooth and slow for equilibrium to be maintained.
The decomposition in b) comes about after \ys{the} careful analysis and decomposition of the forces on a neutrally buoyant particle in~\cite{Harding2019JFM} which, for example, reveals that the centripetal and centrifugal forces are approximately equal and opposite for a neutrally buoyant particle.
For c) we use a novel approximation of inertial lift which preserves the topological structure of the zero level curves of the inertial lift force to ensure an accurate prediction of equilibria points over a wide range of problem parameters.
Put together, these simplifications result in our Zero Level Fit (ZeLF) model, first introduced here, which expedites the calculation of particle trajectories allowing a detailed study of the resulting dynamical system.

Using the ZeLF model we study the dynamics of particle migration in a curved square duct. \ys{We show that there are three regimes, a small $\kappa$} regime in which a small number of stable equilibria exist, \ys{a large $\kappa$} regime in which a stable attracting orbit exists, and the transition between these two regimes. \ys{This intermediate $\kappa$ regime is of particular interest as it has only one focusing point.
For this regime, we investigate the axial distance and time scale required for the focusing to occur, and how these depend on initial particle location in the cross-section.}
While previous studies have only looked at the dynamics within the cross-section, the axial dynamics are extremely important in the context of applications (e.g. cell isolation and separation) in which it is necessary for particles to \ys{be focused by the time they reach} the end of the \ys{duct}.

% CONTENTS OF THE PAPER
% BH: I have filled this out roughly for now, although it should be revised when the paper is more complete.
The paper is organized as follows.
Section~\ref{sec:background} reviews the previous work done on particle flows in curved ducts. 
Section~\ref{sec:model_construction} describes how the ZeLF model for the particle dynamics is constructed. 
In particular, this section details the construction of the different components and how they are ultimately assembled for estimating particle trajectories to quickly and easily study the dynamics. \ys{In addition, the accuracy of the ZeLF model is shown by comparing it with the numerical model of~\cite{Harding2019JFM}.}
\ys{The ZeLF model is then used in} Section~\ref{sec:toy_model_results} \ys{to analyse} the \ys{migration} dynamics of a small particle \ys{for the three regimes of} small, intermediate and large $\kappa$ value. \ys{How these compare with the dynamics predicted by the numerical model of~\cite{Harding2019JFM} is also illustrated.}
%The intermediate $\kappa$ regime is of particular interest as it has only one focusing point.
%For this regime we investigate the axial distance and time scale required for the focusing to occur.
%We also provide a comparison with ZeLF model to the data computed in~\cite{Harding2019JFM}. 
%Finally we make some conclusions.
\ys{Conclusions are given in Section~\ref{sec:Conclusions}.}

\section{Background}\label{sec:background}

\begin{figure}[t]
    \centering
    \begin{subfigure}[b]{0.65\textwidth}
        \centering
        \includegraphics[height = 5cm]{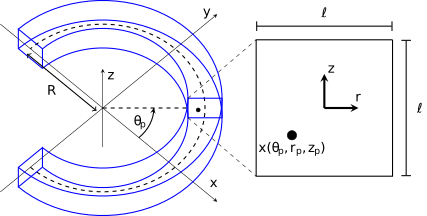} 
        \caption{}
        \label{fig:curved_square_duct}
    \end{subfigure}
    \hfill
    \begin{subfigure}[b]{0.34\textwidth}
        \centering
        \includegraphics[height = 5cm]{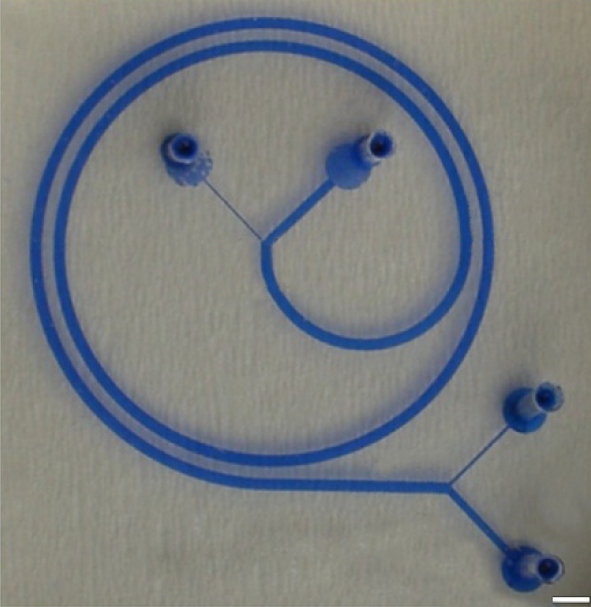}
        \caption{}  
        \label{fig:physical photo of spiral duct}
    \end{subfigure}
\caption{Configurations of curved ducts.
(a) Curved duct with square cross-section containing a spherical particle located at $\mathbf{x}_{p}=\mathbf{x}(\theta_{p},r_{p},z_{p})$. 
The enlarged view of the cross-section around the particle illustrates the origin of the local $r,z$ coordinates at the center of the duct, as first described in \cite{Harding2019JFM}.
The bend radius $R$ is with respect to the center-line of the duct and is quite small here for illustration purposes.
Note that we do not consider the flow near the inlet/outlet.  
\red{(b) Photo of an actual curved microchannel, provided by and reproduced with the permission of the Warkiani Laboratory, University of Technology Sydney, Australia. Notice the bend radius is approximately constant for $7/4$ turns. The scale bar on the bottom right is 2cm.}}
        \label{figs:curved_square_duct}

\end{figure}

The general setup of our curved square duct is depicted in Figure~\ref{figs:curved_square_duct}.
The \ys{horizontal and vertical} coordinates within the duct cross-section are $(r,z)\in[-\ell/2,\ell/2]\times[-\ell/2,\ell/2]$ where $\ell$ is the side length of the square cross-section.
The cross-sectional coordinates map to points in \ys{the} three dimensional duct via
\begin{equation}
\mathbf{x}(\theta,r,z) = (R+r)\cos(\theta)\mathbf{i}+(R+r)\sin(\theta)\mathbf{j}+z\mathbf{k} \,,
\end{equation}
where \ys{$\theta$ is angular distance along the duct and} $R$ is the bend radius of the duct (measured to the center of the cross-section).
\ys{The dimensionless parameter} $\epsilon:=\ell/(2R)$ is used to characterise the bend radius relative to \bh{half} the cross-section \ys{height}.
%We later use non-dimensionalized coordinates $(\tilde{s},\tilde{z})=(2s/\ell,2z/\ell)\in[-1,1]\times[-1,1]$ to describe coordinates within the duct cross-section. 

Let $\bar{p},\bar{\mathbf{u}}$ describe the pressure and velocity fields, respectively, of a steady pressure driven flow through the curved square duct in the absence of any particles. %, modeled with the Navier--Stokes equations.
The fluid is assumed to be incompressible with uniform density $\rho$ and uniform viscosity $\mu$.
For convenience we \bh{separate} $\bar{\mathbf{u}}$ into its axial component $\bar{\mathbf{u}}_{a}$ and secondary component $\bar{\mathbf{u}}_{s}$, specifically 
\begin{align*}
\bar{\mathbf{u}}_{a} &:= (\bar{\mathbf{u}}\cdot \mathbf{e}_{\theta})\mathbf{e}_{\theta} \,, \\
\bar{\mathbf{u}}_{s} &:= \bar{\mathbf{u}}-\bar{\mathbf{u}}_{a} \,.
\end{align*}
The maximum of $\bar{\mathbf{u}}_{a}$ is taken to be \ys{the} characteristic flow velocity and is denoted by $U_{m}$.
It is assumed that the Dean number $K=\Rey_{c}^{2}\epsilon$, \ys{where} \red{$\epsilon:=\ell/(2R)$ (as defined earlier) and }$\Rey_{c}=(\rho/\mu)U_{m}\ell/2$ is the channel Reynolds number, is small enough that the inertia of the fluid flow through the curved duct does not perturb the axial velocity component significantly from the Poiseuille flow obtained in a straight duct.
The secondary flow $\bar{\mathbf{u}}_{s}$, consisting of two counter rotating vortices in the cross-sectional plane, scales with $U_{m}\sqrt{K\epsilon}$~\cite{Harding2018ANZIAMJ}.

A spherical particle with radius $a<\ell/2$ is then suspended in the fluid flow resulting in the new pressure and velocity fields $p,\mathbf{u}$, respectively.
The location of the particle is described by the location of its center $\mathbf{x}_{p}=\mathbf{x}(\theta_{p},r_{p},z_{p})$.
\red{We assume the particle is free to spin, which %is important to consider when calculating 
features in the calculation of the forces that %which 
influence its %the 
motion %of a particle~
\cite{Harding2019JFM}. However, in this study we do %need 
not track the particle spin %in this study
as our aim is to produce a simplified model of the particle's position.}
% \red{While the spin affects the flow as is described in ~\cite{}, as we will see later, 
% we do n
% particle is free to spin as it migrates through the duct and affects the flow.
% However due to its spherical symmetry it is not necessary to track the rotation itself.}
% The particle is free to spin as it migrates through the duct (although due to its spherical symmetry it is not necessary to track \ys{the rotation}).
The flow fields $p,\mathbf{u}$ are non-steady due to the motion of the spherical particle suspended in it.
\ys{However, as in \cite{Harding2019JFM}, we move to the reference frame rotating with the cross-section containing the particle center, in which the angular coordinate is $\theta'=\theta-\theta_p$ and the radial and vertical coordinates remain unchanged. In this rotating reference frame, the fluid motion may be taken as steady}
and a disturbance flow $q',\mathbf{v}'$ is introduced which describes the difference between $\bar{p},\bar{\mathbf{u}}$ and $p,\mathbf{u}$ in that frame.
The force on the particle can be decomposed into three components: a gravity/buoyancy balance $\mathbf{F}_{g}$, a centrifugal/centripetal force balance $\mathbf{F}_{c}$ and a remaining hydrodynamic component $\mathbf{F}_{nb}$.
There is an analogous decomposition of the torque which we do not describe in detail.

% BH: I've moved this from the introduction
A recent paper explored the effects of non-neutral particle buoyancy for curved ducts having a rectangular cross-section~\cite{Harding2020JFM}. 
Perturbations due to non-neutral buoyancy were found to be small for Froude numbers larger than $3$, $\mathrm{Fr}^{2}=U_{m}^{2}a/(g\ell^{2})$, and particle density $\rho_{p}$ satisfying $|\rho_{p}-\rho|\leq\rho/2$. 
Since this is the case for typical applications of inertial microfluidics, for simplicity, we \ys{here restrict attention to} neutrally buoyant particles.

For a neutrally buoyant particle \ys{(for which $\mathbf{F}_{g}=0$} \bh{and $\mathbf{F}_{c}\approx0$ once axial equilibrium is achieved)} only the hydrodynamic \ys{force} component remains, \ys{namely}
$$
\mathbf{F}_{nb} = \int_{|\mathbf{x}'-\mathbf{x}_{p}^{\prime}|=a}\mathbf{n}\cdot\left(-q'\mathbf{I}+\mu\left(\nabla'v'+\nabla'v^{\prime\,T}\right)\right)\,dS'\,,
$$
where $|\mathbf{x}'-\mathbf{x}_{p}^{\prime}|=a$ is the surface of the particle, \ys{primes denote variables in the rotating reference frame,} %($\mathbf{x}'$ and $\mathbf{x}_{p}^{\prime}$ being coordinates in the appropriate rotating frame) 
and $\mathbf{n}$ is the outward pointing normal.
Upon non-dimensionalizing for a viscous flow, using the characteristic length $a$ and velocity $U=U_{m}a/\ell$, one may perform a perturbation expansion of $q',\mathbf{v}'$ with respect to the particle Reynolds number $\Rey_{p}=(\rho/\mu)U_{m}a^{2}/\ell$, \ys{which is assumed to be} sufficiently small.
\red{Notice that particles are expected to quickly approach terminal velocity as determined by Stokes' drag law and, thereafter, $\Rey_{p}$ is the effective Stokes number during particle migration.} 
%\red{Notice that $\Rey_{p}$ is effectively the Stokes number. Therefore, the particles while quickly approach the terminal velocity determined by Stokes' drag law.} 
For convenience \ys{and since $\bar{\mathbf{u}}_{s}$ scales with $U_{m}\sqrt{K\epsilon}=\Rey_{p}U\kappa$, where 
\begin{equation}
    \kappa=\ell^{4}/(4a^{3}R), \label{kappa_def}
\end{equation}
the} secondary component of the background flow velocity is considered to be $\mathcal{O}(\Rey_{p})$.
Consequently, the leading order component of $\mathbf{F}_{nb}$ describes the primary force balance governing the axial velocity of the particle \ys{$u_{p}=d\theta_p/dt$} (and analogously the leading order component of the torque describes the primary balance governing its spin $\mathbf{\Omega}_{p}$).
%Upon fixing $u_{p,\theta}$ and $\mathbf{\Omega}_{p}$ to zero the leading order components, the first order component of $\mathbf{F}_{nb}$ provides a balance of the forces governing particle migration within the cross-section.
%Specifically this first order component balances the inertial lift $\mathbf{L}$ and secondary flow drag $\mathbf{D}$.

The first order component of $\mathbf{F}_{nb}$ describes the forces governing particle migration within the cross-section, specifically the inertial lift $\mathbf{L}$ and secondary flow drag $\mathbf{D}$.
The perturbation analysis \ys{yields}~\cite{Hood2015JFM} %suggests that 
\begin{equation}
\mathbf{L}\propto \rho U_{m}^{2}a^{4}/\ell^{2}=\mu a U \Rey_{p} 
\end{equation}
\ys{and,} recalling that $\bar{\mathbf{u}}_{s}\propto \Rey_{p}U\kappa$ \ys{and} using \ys{the} Stokes drag law to approximate the secondary drag force, it follows that 
\begin{equation}
\mathbf{D}\approx 6\pi \mu a \bar{\mathbf{u}}_{s}\propto \mu a U \Rey_{p} \kappa \,.
\end{equation}
Therefore, one expects \ys{$\mathbf{D}\propto \kappa\mathbf{L}$.} %$\mathbf{D}$ to scale as $\kappa$ relative to $\mathbf{L}$.

Computational approximations of $\mathbf{L}$ and $\mathbf{D}$ \ys{have been previously} obtained over several \ys{cross-sectional} shapes (square, rectangular and trapezoidal), \ys{in each case} for a \ys{number} of particle \ys{sizes} and duct bend radii, \ys{including for straight ducts ($R\rightarrow\infty$)}~\cite{Harding2019JFM}.
%Although \ys{$u_{p}$} was also computed in the process it was \ys{not reported} given the focus \ys{in that paper} on studying the cross-sectional dynamics. 
A key factor in determining the stability of equilibria and the existence of slow manifolds is the \ys{intersection} of zero level curves of the \ys{$r$ and $z$} components of the net migration force $\mathbf{L}+\mathbf{D}$.
\ys{Using these, a number of significantly different types of migration dynamics were identified. For rectangular cross-sectional shape in particular, but also for trapezoidal cross-sections, these were found to depend on the value of $\kappa$.} %illustrates many of the significant changes in migration dynamics.
%This was observed in case of the ducts having a rectangular cross-section (with aspect ratio 2 and 4 in particular) as $\kappa$ approximately characterised the location of stable focusing equilibria.
%

\ys{Ducts with square cross-sections were less studied in~\cite{Harding2019JFM} but three different types of migration dynamics were identified, characterized by four stable equilibria, a single stable equilibrium, and a pair of stable limit cycles. In this paper, we undertake a detailed examination of migration dynamics in ducts with square cross-sectional shapes.} %The square cross-section also illustrated some interesting migration dynamics, focusing to a single point and limit cycles, which we explore in great detail in this paper.
%
%While a large amount of simulation data describing particle migration in curved ducts was generated by Harding et. al \cite{Harding2019JFM}, 
\ys{For this purpose,} it is desirable to construct a \ys{simpler} model, which we call the Zero Level Fit (ZeLF) model, which is more tractable for studying the dynamical system of the particle motion in depth.

%Considering a small particle in a duct with reasonably large bend radius Harding~\cite{Harding2018AFMC} considered a simpler approximation cross-sectional trajectories.
%In particular, it was assumed only $\mathbf{L}$ and $\mathbf{D}$ drive cross-sectional migration and effects of acceleration (including the added mass force) can be neglected.

\section{Constructing the ZeLF model}\label{sec:model_construction}
%Need to condense discussion below and remove duplication

\ys{The inertial lift $\mathbf{L}$ experienced by the particle is primarily due to the \bh{disturbance of the axial} flow along the duct.
We assume that $\mathbf{L}$ is well approximated by that for the case of flow through a duct with the same cross-section in the limit $R\rightarrow\infty$ ($\epsilon\rightarrow 0$). 
To this, we add the effect of drag force $\mathbf{D}$ approximated by the Stokes drag law applied to the secondary velocity field $\bar{\mathbf{u}}_{s}$ describing the flow vortices in the cross-section that are due to the curvature of the duct also obtained for the limit $\epsilon\rightarrow 0$.
Specifically, we compute 
$\hat{\mathbf{u}}_s=\lim_{R\to\infty}\frac{\bar{\mathbf{u}}_{s}}{U_m\sqrt{K\epsilon}}$
and then subsequently approximate $\bar{\mathbf{u}}_{s}$ as
$\bar{\mathbf{u}}_{s}\approx U_m\sqrt{K\epsilon}\hat{\mathbf{u}}_s$.
A similar model was briefly explored in \cite{Harding2018AFMC}, focusing on a duct having a rectangular cross-section and using numerical simulation data to compute the forces; for a sufficiently small ratio of \bh{duct height $\ell$} to bend radius $R$, the predicted net force driving particle migration was found to be similar to that of the more complex model~\cite{Harding2019JFM} summarized in the previous section. 

However, in contrast to the model of \cite{Harding2018AFMC}, here we fit simple model functions to simulation data for a \bh{small particle suspended in flow through a curved} duct having a square cross-section and use these functions to \bh{estimate} the force components.} %Firstly we construct
%Our ZeLF model is a further simplification of this approximation.
%This consists of (a) constructing 
%a polynomial approximation of $\bar{\mathbf{u}}_{s}$ from which to compute $\mathbf{D}$. Secondly we construct}
%, and (b) constructing 
%a simple formula approximating $\mathbf{L}$ %which the correct migration dynamics
%\ys{for} a straight square duct.
%The latter consists of first constructing a (bivariate) polynomial that captures the zero level curves of the inertial lift force before fitting an exponential term to capture/correct the magnitude elsewhere.
\bh{Our model functions preserve} the topology of the zero level sets \ys{for the inertial lift and drag} allowing us to retain the correct migration dynamics for ducts with large bend radius.
Additionally, the model smooths over some of the numerical noise/error which is present in the simulation data.

A weakness of this modeling approach is that it may not be appropriate for ducts having a smaller bend radius where the curved geometry of the duct has a noticeable influence on both the inertial lift force and the secondary drag force.
\ys{As the bend radius increases,} the influence of the curved geometry on both of these components decays.
Therefore our modeling of both the inertial lift force and secondary drag forces are expected to be most accurate for ducts having a large bend radius \ys{compared with the cross-sectional width}.

%Rather than continue to neglect the axial component of the motion, herein we re-couple this with the cross-sectional dynamics, in a simplified manner.
Understanding the axial distance traveled by a particle during migration is crucial to the design of devices \ys{for particle focusing and separation. Hence, herein we couple this to the cross-sectional dynamics in a simplified manner}.
We take the axial velocity of the particle to always be equal to the terminal velocity if it were to remain fixed at its current cross-section location. 
This is a reasonable assumption because the axial velocity is expected to reach equilibrium on a much faster time scale than that of the cross-sectional migration.
Further, since the axial velocity of the particle (at equilibrium) is similar to that of the background flow (there is a small `slip' but it is order $(a/\ell)^{2}$) then this may be approximated by $\bar{\mathbf{u}}_{a}$\bh{$(r_p,z_p)$}.
Similar to $\bar{\mathbf{u}}_{s}$, we approximate $\bar{\mathbf{u}}_{a}$ with a \ys{simple model function for the limit $\epsilon\rightarrow 0$.} %(bivariate) polynomial.

In the remainder of the paper we describe our model non-dimensionalising the cross-sectional coordinates according to
\begin{equation}
(r,z) = \left(\frac{\ell}{2}\tilde{r},\frac{\ell}{2}\tilde{z}\right) \,.
\end{equation}
This is most convenient because the rescaled cross-sectional domain $(\tilde{r},\tilde{z})\in[-1,1]\times[-1,1]$ is independent of any physical parameters (in contrast \ys{to} scaling with respect to the particle radius $a$ \ys{for which} the non-dimensionalized duct dimensions \ys{are} $\ell/a$).
%For brevity the tilde above $r,z$ will be dropped.

%\section{Constructing the ZeLF model}\label{sec:model_construction}

% BH: High level description of the model
%Our modeling of the forces on the particle effectively assumes that the particle experiences an inertial lift force as if the duct is straight, then adding to this the effect of drag force from the secondary flow vortices that occur in a curved duct, effectively treating this as an additional perturbation to the migration behavior.
%We assume that the distance between each particle is large compared to the duct size $l$, and ignore any particle-particle interactions.
%A similar model was briefly explored in \cite{Harding2018AFMC} albeit using numerical simulation data directly, and also focusing on a duct having a rectangular cross-section.
%In contrast, here we fit simple model functions to simulation data for a duct having a square cross-section and then go on to study the particle migration dynamics in much greater detail.

%
The following subsections outline the construction of the different model components %in more detail 
before describing how they are assembled for the computation of particle trajectories.
%Given one of our aims is to re-couple the axial motion of the particle with the cross-sectional dynamics, we begin by fitting a simple polynomial approximation to the axial velocity profile of a slow and steady flow through the duct in the absence of a particle.
%This will provide a reasonable approximation to the particle's axial velocity.
%We then fit a similar polynomial approximation to the streamfunction describing the secondary motion of the flow.
%By taking the derivatives of this and applying Stokes' drag law we obtain a simple approximation of the secondary flow drag on the particle.
%The development of an approximation of the inertial lift force is then described.
%Finally the components are assembled into a first order system of ordinary differential equations which approximate the dynamics of a particle suspended in the flow through a curved duct.

\subsection{Modeling the inertial lift force}\label{Modeling the inertial lift forces} % BH

\ys{We first give the model for the inertial lift acting on a particle and then discuss its derivation.} The dimensionless \ys{component of lift} in the $r$ direction within the cross-section, \ys{for a particle with position $(\tilde r_p,\tilde z_p)=(\tilde r,\tilde z)$, is approximated by}
%%% The following model is with the parameters a=0.05,N=2
%$$
%\hat{L}_{s}(s,z) = s(1-12.7s^{6}-24.8z^{6})
%e^{2.95-2.58s^{2}-1.91z^{2}+1.44s^{4}-0.7s^{2}z^{2}-1.63z^{4}}
%$$
%%% The following model is with the parameters a=0.05,N=3
\begin{multline}\label{eqn:Lr_model}
\hat{L}_{r}(\tilde{r},\tilde{z}) = \tilde{r}\left(1-12.7\tilde{r}^{6}-24.8\tilde{z}^{6}\right)
\exp\left(2.95-1.43\tilde{r}^{2}-4.23\tilde{z}^{2}-1.98\tilde{r}^{4}\right.\\ \left.+5.28\tilde{r}^{2}\tilde{z}^{2}-1.10\tilde{z}^{4}+2.35\tilde{r}^{6}-1.16\tilde{r}^{4}\tilde{z}^{2}-7.16\tilde{r}^{2}\tilde{z}^{4}+3.51\tilde{z}^{6}\right).
\end{multline}
\ys{Similarly, because of the expected symmetry for a straight duct with square cross-section,
the component in the $z$ direction is} approximated as 
\begin{equation*}
\hat{L}_{z}(\tilde{r},\tilde{z})=\hat{L}_{r}(\tilde{z},\tilde{r}).   
\end{equation*} 
In the dimensional setting these two inertial lift force components are
\begin{equation*}
L_{r},L_{z} = \rho U_{m}^{2}\frac{a^{4}}{\ell^{2}}\hat{L}_{r},\rho U_{m}^{2}\frac{a^{4}}{\ell^{2}}\hat{L}_{z} \,.
\end{equation*}
% This scaling of the inertial lift force \ys{was} derived \ys{using} the particle radius $a$ as the characteristic length scale, and $U_{m}a/\ell$ as the characteristic velocity scale. \ys{Comment: I do not see the need to state how the scaling was derived and, so, would remove this last sentence.}\kh{KH: this should be addressed}
%This velocity scale, sometimes referred to as the particle shear velocity, approximates the magnitude of the change in the background fluid velocity across the diameter of the particle.

\begin{figure}[t]
    \centering
    \begin{subfigure}[b]{0.49\textwidth}
        \centering
        \includegraphics{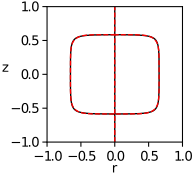}
        \caption{The zero level set curve of $L_{r}$}\label{fig:Ls_zls}
    \end{subfigure}
    \hfill
    \begin{subfigure}[b]{0.49\textwidth}
        \centering
        \includegraphics{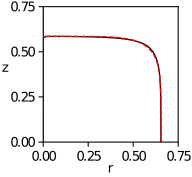}
        \caption{Zoom of the zero level set curve of $L_{r}$}\label{fig:Ls_zls_zoom}
    \end{subfigure}
    \caption{Fit of the zero level set curve of \ys{$f(\tilde r,\tilde z)=\tilde{r}(1-12.7\tilde{r}^{6}-24.8\tilde{z}^{6})$  (in red) with that} of $\tilde L_{r}$ \ys{from} finite element computations from~\cite{Harding2019JFM} \ys{(in black)}. Figure (a) shows the two over the entire cross-section whereas figure (b) zooms into a portion of the upper right quadrant. The difference between the two is difficult to discern at both scales.}\label{fig:Ls_zls_model}
\end{figure}

\begin{figure}[t]
    \centering
    \begin{subfigure}[b]{0.49\textwidth}
        \centering
        \includegraphics{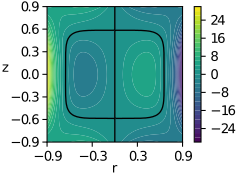}
        \caption{Inertial lift model $\hat{L}_{r}$}\label{fig:Ls}
    \end{subfigure}
    \hfill
    \begin{subfigure}[b]{0.49\textwidth}
        \centering
        \includegraphics{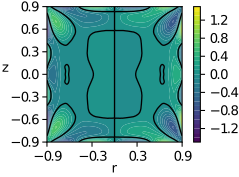}
        \caption{Error of $\hat{L}_{r}$}\label{fig:Ls_error}
    \end{subfigure}
    \caption{\ys{Model of the inertial lift force $\hat{L}_{r}(\tilde{r},\tilde{z})$: (a) $\hat{L}_{r}$ over the cross-section excluding a small region around the walls; (b) the difference between the model and results computed from finite element solutions~\cite{Harding2019JFM}.}}\label{fig:Ls_model}
\end{figure}

The inertial lift force model was constructed as follows.
First the \ys{factor preceding the exponential in $\hat L_r$, $f(\tilde r,\tilde z)=\tilde{r}(1-12.7\tilde{r}^{6}-24.8\tilde{z}^{6})$,} was determined \ys{by trial and error to give a good match between its} zero level contour \ys{and} the zero level contour of an interpolation of $L_{r}$ data computed via numerical simulations \ys{for} a small particle (specifically with $a/\ell=1/20$) in a straight duct as described in~\cite{Harding2019JFM}; see Figure~\ref{fig:Ls_zls_model}.
\ys{This ensures the prediction of} the correct location and stability of equilibria for a straight duct. 
%This initial fitting of the zero level curve inspired the naming of the model: the zero level fit model (or ZeLF).
The exponential factor, with exponent consisting of a polynomial in $\tilde{r},\tilde{z}$, \ys{was then} added to improve the global accuracy of the model in a way that does not modify the zero level contours.
The coefficients of the polynomial (in the exponent) \ys{were} obtained via a constrained least squares fitting to \ys{the} interpolation of the $L_{r}$ data from the numerical simulations.
There is a classical trade-off between accuracy and simplicity of the model in determining a suitable degree of the polynomial within the exponent. 
\ys{Compared with the simulation data, the} specific model \eqref{eqn:Lr_model} achieves a $L_2$ relative error of $3.8\%$, see Figure~\ref{fig:Ls_model}. %(compared with an interpolant of the simulation data). % Note: the comparison is over the domain of the data and excludes a small region adjacent to the walls, but this is a subtle point I don't think we need to dwell on. % Further: The error of the a=0.05,N=2 model is $8.9\%$.

\subsection{Modeling the secondary drag force} \label{Modeling the secondary drag force}
% BH: this could be simplified, e.g. we possibly don't need all of these details

The secondary drag force on the particle \ys{is} approximated by combining Stokes' drag law with the velocity of the secondary component of the fluid flow through a curved duct \ys{in the limit $\epsilon\rightarrow 0$}.
The secondary component consists of two counter-rotating vortices which are orthogonal to the main direction of flow.
For a slow laminar flow through a curved duct, it can be shown that the velocity of the secondary component scales as \bh{$U_{m}\sqrt{K\epsilon}=\epsilon\mathrm{Re}_{c}U_{m}$} \cite{Harding2018ANZIAMJ,Harding2019JFM}. %$\epsilon\mathrm{Re}_{c}U_{m}$ where $\epsilon=\ell/(2R)$ and $\mathrm{Re}_{c}=(\rho/\mu) U_{m}(\ell/2)$ \cite{Harding2018ANZIAMJ,Harding2019JFM}. 
The two velocity components can be described via a stream-function $\Phi(\tilde{r},\tilde{z})$, specifically with $-\partial\Phi/\partial \tilde{z}$ and $\partial\Phi/\partial \tilde{r}$ describing the velocity in the $\tilde r$ and $\tilde z$ directions, respectively.

The \bh{fundamental scale and topology} of $\Phi$ % for a slow laminar flow through a curved duct 
\ys{is} approximated as
$$
\Phi_{0}(\tilde{r},\tilde{z}) = -0.01591\epsilon\mathrm{Re}_{c}U_{m}(1-\tilde{r}^{2})^{2}\tilde{z}(1-\tilde{z}^{2})^{2} \,.
$$
This approximation ensures that both $-\partial\Phi/\partial \tilde{z}$ and $\partial\Phi/\partial \tilde{r}$ are zero on the duct walls, \ys{describes} two counter rotating vortices with the correct orientation, \ys{has} the desired odd symmetry with respect to $\tilde z$, and \ys{has} even symmetry with respect to $\tilde r$ \ys{as required} in the limit $\epsilon\rightarrow 0$. 
The factor $0.01591$ was determined to fit the velocity fields obtained from a finite difference solution of the Navier--Stokes equations governing the background flow in the limit \ys{$\epsilon\rightarrow 0$} and \ys{at} small flow rate.

\begin{figure}[t]
    \centering
    \begin{subfigure}[b]{0.49\textwidth}
        \centering
        \includegraphics{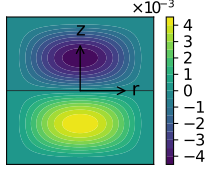}
        \caption{Stream-function model $\Phi_{1}$}\label{fig:Phi1_model}
    \end{subfigure}
    \hfill
    \begin{subfigure}[b]{0.49\textwidth}
        \centering
        \includegraphics{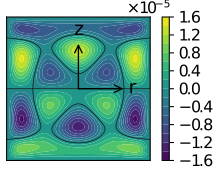}
        \caption{Error of $\Phi_{1}$}\label{fig:Phi1_model_error}
    \end{subfigure}
    \caption{\ys{Model of the \ys{secondary vortices:} (a) streamfunction $\Phi_{1}(r,z)$ and (b) the difference between $\Phi_{1}(r,z)$ and the streamfunction from a finite difference computation over the cross-section in the limit $\epsilon\rightarrow 0$.}}\label{fig:Phi_model}
\end{figure}

%As with axial velocity, t
\ys{The} accuracy of the secondary velocity approximation can be improved with the addition of further terms.
%Analogous to the modeling of $u_{N}$, w
\ys{We have performed} an $L_{2}$ fit of the model 
\begin{equation*}
\Phi_{N}(\tilde{r},\tilde{z})=\Phi_{0}(\tilde{r},\tilde{z})\sum_{n=0}^{N}\sum_{i=0}^{n}\beta_{2i,2(n-i)}\tilde{r}^{2i}\tilde{z}^{2(n-i)},
\end{equation*}
\ys{for $N\in\mathbb{N}$,} to the stream-function computed from \ys{the above mentioned} finite difference solution of the Navier--Stokes equations. % (in a limit of large bend radius and low flow rate).
\ys{Fits were determined} for several $N$ but $N=1$ was  \ys{found} to be sufficiently accurate for our study.
In particular, we obtained 
\begin{align*}
%\Phi_{1}(s,z) &= \frac{\Phi_{0}(s,z)}{0.01591}(0.01564+0.003640s^{2}-0.0002831z^{2}) \,,\\
%\Phi_{2}(s,z) &= \frac{\Phi_{0}(s,z)}{0.01591}\left(\begin{array}{l}0.01560+0.003203s^{2}+0.0002693z^{2}\\+0.001540s^{4}-0.0007749s^{2}z^{2}-0.0008193z^{4}\end{array}\right) \,.
\Phi_{1}(\tilde{r},\tilde{z}) &= \Phi_{0}(\tilde{r},\tilde{z})(0.9833+0.2289\tilde{r}^{2}-0.0178\tilde{z}^{2}) 
% \,,\\
% \Phi_{2}(r,z) &= \Phi_{0}(r,z)(0.9807+0.2014r^{2}+0.0169z^{2}+0.0968r^{4}-0.0487r^{2}z^{2}-0.0516z^{4}) \,,
\end{align*}
where the coefficients have been rounded to four decimal places.
%These attain all of the feature of $\Phi_{0}$ described but have better accuracy with respect to a computed solution. 
A plot of the approximation $\Phi_{1}$ is shown in Figure~\ref{fig:Phi_model} alongside a plot of the error \ys{with respect to the finite-difference solution}.
The relative error of $\Phi_{1}$ is found to be $0.3\%$ making it sufficiently accurate for the study of dynamics herein.

For a small spherical particle, the drag force within the cross-sectional plane can be \bh{estimated} using Stokes' drag law in conjunction with the velocities obtained from the stream-function $\Phi_{1}$.
Specifically, for a small particle suspended in the flow and which is not moving with respect to the \ys{$\tilde r,\tilde z$} coordinates, \ys{one} can use the approximation
$$
D_{r},D_{z} = -6\pi\mu a\frac{\partial\Phi_{1}}{\partial \tilde{z}},6\pi\mu a\frac{\partial\Phi_{1}}{\partial \tilde{r}}
$$
for the \ys{radial and vertical} components of the drag force respectively.
Taking the scaling of \ys{$\Phi_1$} into account, it is reasonable to non-dimensionalize the secondary drag force according to
\begin{align*}
D_{r},\,D_{z} 
&= \mu a \epsilon \Re_{c} U_{m} \tilde{D}_{r},\, \mu a \epsilon \Re_{c} U_{m} \tilde{D}_{z} \\
&= \rho U_{m}^{2} \frac{a\ell^{2}}{4R} \tilde{D}_{r},\, \rho U_{m}^{2} \frac{a\ell^{2}}{4R} \tilde{D}_{z} \,.
\end{align*}

It is important to note that this approximation will not be accurate when the particle is very close to a wall, or for larger particles.
The true drag coefficient is larger for increasing particle size, and additionally the drag coefficient increases when the particle approaches a wall.
%However, since the same drag coefficient is used in the final calculation of migration velocity from the net migration force, then the specific coefficient doesn't really matter for our model of the secondary drag force.
%Thus, we stick with Stokes' drag law for simplicity rather than choosing to add corrections to the drag coefficient. \ys{Not sure if this para should be here or later.}
\bh{Since our interest is primarily smaller particles and their dynamics away from the walls we stick with Stokes' drag law to maintain the simplicity of the model.}

\subsection{Modeling the axial velocity}

The particle travels with an axial velocity which is close to that of the background fluid flow.
For a slow laminar fluid flow through a curved square duct (with no particles), the axial velocity field is quite close to Poiseuille flow \ys{in a straight square duct}, specifically within $O(\epsilon)$.
With $U_{m}$ denoting the maximum axial velocity, then a simple approximation of the dimensionless Poiseuille flow is given by
\begin{equation}\label{eqn:u0}
u_{0}(\tilde{r},\tilde{z}) = U_{m}(1-\tilde{r}^2)(1-\tilde{z}^2) \,.
\end{equation}
This approximation attains the expected maximum, satisfies no-slip boundary conditions on the walls, and has the even symmetry with respect to $r$ and $z$ that is expected for \ys{this} flow.
%\bh{(Although even symmetry with respect to $r$ is, in reality, not correct for flow through a curved duct, the non-symmetric component is small at low flow rates, and additionally decays with $\epsilon$, so it is reasonable to neglect in the current study.)}

\begin{figure}[t]
    \centering
    \begin{subfigure}[b]{0.49\textwidth}
        \centering
        \includegraphics{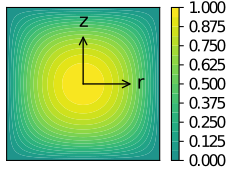}
        \caption{Axial velocity model $u_{1}$}\label{fig:u1_model}
    \end{subfigure}
    \hfill
    \begin{subfigure}[b]{0.49\textwidth}
        \centering
        \includegraphics{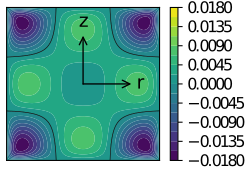}
        \caption{Error of $u_{1}$}\label{fig:u1_model_error}
    \end{subfigure}
    \caption{(a) Model \ys{function for} the axial velocity $u_{1}(\tilde{r},\tilde{z})/U_m$ and (b) the difference between this and a finite difference computation of Poiseuille flow through a straight square duct with unit wall length.}\label{fig:u_model}
\end{figure}

The accuracy of the simple approximation \eqref{eqn:u0} degrades away from the wall and the center, and can be improved by the addition of further terms \ys{in a similar manner to that used above for the streamfunction}.
%There are several ways the velocity field can be modeled, see for example \cite{HardingStokes2018POF} and \cite{Harding2018ANZIAMJ} among others.
\ys{Thus,} we perform a simple $L_{2}$ fit of the model
\begin{equation}\label{eqn:u_N}
u_{N}(\tilde{r},\tilde{z})=u_{0}(\tilde{r},\tilde{z})\left(1+\sum_{n=1}^{N}\sum_{i=0}^{n}\alpha_{2i,2(n-i)}\tilde{r}^{2i}\tilde{z}^{2(n-i)}\right),
\end{equation}
for a given $N\in\mathbb{N}$, with an axial velocity field computed from a finite difference solution of \bh{Poiseuille flow} \ys{in a straight duct}.
\bh{The general form \eqref{eqn:u_N}} retains all of the features of $u_{0}$ described above but provides a better approximation for increasing $N$.
We constructed approximations for several $N$ but \ys{here too} found that $N=1$ was sufficiently accurate for our study.
Specifically, we obtained
\begin{equation*}
    u_{1}(\tilde{r},\tilde{z}) = u_{0}(\tilde{r},\tilde{z})\left(1+0.1818(\tilde{r}^{2}+\tilde{z}^{2})\right),
\end{equation*}
% \,,\\
% u_{2}(r,z) &= u_{0}(r,z)\left(1+0.1410(r^{2}+z^{2})+0.3742r^{2}z^{2}+0.0082(r^{4}+z^{4})\right) \,,
where the coefficient has been rounded to four decimal places. 
A plot of the approximation $u_{1}$ is shown in Figure~\ref{fig:u_model} alongside a plot of the error $u_{1}-\ys{u_a}$ \ys{(both scaled with $U_m$)} where $\ys{u_a=}\bh{\lim_{R\rightarrow\infty}}\bar{\mathbf{u}}_{a}\cdot\mathbf{e}_{\theta}$, \ys{i.e. the axial flow through a straight square duct}.
The relative error \ys{$\|u_1-u_a\|_2/\|u_a\|_2$} %$\|u_{1}-\bh{\lim_{R\rightarrow\infty}}\bar{\mathbf{u}}_{a}\cdot\mathbf{e}_{\theta}/U_{m}\|_{2}/\|\bar{\mathbf{u}}_{a}/U_{m}\|_{2}$
is found to be $1.1\%$ making $u_{1}$ sufficiently accurate for the study of dynamics herein.
\bh{The terminal velocity of a particle is thus approximated as 
\begin{equation*}
u_{p} \approx u_{1}(r_{p},z_{p}) \,.
\end{equation*}}

\subsection{Putting the model together} \label{sec: putting the model together}

We now approximate the net force on the particle in the $r,z$ directions as 
\begin{equation*}
F_{r}=L_{r}+D_{r} \,,\quad F_{z}=L_{z}+D_{z} \,,
\end{equation*}
respectively.
If we non-dimensionalize $F_{r},F_{z}$ with the same scale as $L_{r},L_{z}$, that is
\begin{equation*}
F_{r}=\rho U_{m}^{2}a^{4}/\ell^{2}\hat{F}_{r} \,,\quad F_{z}=\rho U_{m}^{2}a^{4}/\ell^{2}\hat{F}_{z} \,,
\end{equation*}
then we obtain
\begin{equation*}
\hat{F}_{r}=\hat{L}_{r}+\kappa \tilde{D}_{r} \,,\quad \hat{F}_{z}=\hat{L}_{z}+\kappa \tilde{D}_{z} \,,
\end{equation*}
\ys{with $\kappa=\ell^{4}/(4a^{3}R)$ as defined in \eqref{kappa_def}. This highlights the fact that $\kappa$}
 describes the magnitude of the secondary drag force relative to the inertial lift force.

\ys{Using} Stokes' drag law, the terminal velocity of a small particle due to the net migration force is
$$
v_{r} = \frac{F_{r}}{6\pi\mu a} = \frac{L_{r}+D_{r}}{6\pi\mu a} \,,\quad v_{z} = \frac{F_{z}}{6\pi\mu a} = \frac{L_{z}+D_{z}}{6\pi\mu a} \,.
$$
In this study, it will be convenient to non-dimensionalize these velocities according to the secondary fluid velocity scale (rather than a scaling based on the inertial lift force).
This is because we generally expect the secondary flow to be the dominant effect for a small particle and the inertial lift force can be viewed as a perturbation to this.
In particular, we introduce
$$
v_{r} = \epsilon\Rey_{c}U_{m}\tilde{v}_{r} \,,\quad v_{z} =\epsilon\Rey_{c}U_{m}\tilde{v}_{z} \,.
$$
Consequently, \ys{we} express $\tilde{v}_{r},\tilde{v}_{z}$ as
$$
\tilde{v}_{r} = \frac{1}{6\pi}\left(\frac{\hat{L}_{r}}{\kappa}+\tilde{D}_{r}\right)\,, \quad \tilde{v}_{z} = \frac{1}{6\pi}\left(\frac{\hat{L}_{z}}{\kappa}+\tilde{D}_{z}\right) \,.
$$

%Letting $r_{p},z_{p}=(\ell/2)\tilde{r}_{p},(\ell/2)\tilde{z}_{p}$ be the coordinates of the particle center in the cross-section, 
\ys{Then, the trajectory of a particle with center $(\tilde r_p,\tilde z_p)$} is modeled via the first order system of ordinary differential equations
\begin{subequations}\label{eqn:migration_ode_alt}\begin{align}
\frac{d\tilde{r}_{p}}{d\tilde{t}} &= \tilde{v}_{r}\left(\tilde{r}_{p},\tilde{z}_{p}\right) \,, \\
\frac{d\tilde{z}_{p}}{d\tilde{t}} &= \tilde{v}_{z}\left(\tilde{r}_{p},\tilde{z}_{p}\right) \,, 
\end{align}\end{subequations}
where \ys{$\tilde t$ is dimensionless time, related to physical time $t$ by} $t=R\tilde{t}/(U_{m}\mathrm{Re}_{c})$. % is the non-dimensionalisation of the time scale.

%As an aside, we can perform a simple dimensional analysis to estimate the time scale $T_{m}$ required for inertial migration to occur due to the inertial lift force.
%\ys{Given the migration velocity is proportional} to $\Rey_{p}U_{m}a/\ell$ \ys{(I'm confused by this given we've just seen $v_r,v_z\sim \epsilon Re_cU_m= Re_pU_m\kappa a/\ell$)} and an anticipated migration distance proportional to $\ell$, then we can expect this to occur over 
%\begin{equation*}
%T_{m} \propto \frac{\ell}{\Rey_{p}U_{m}a/\ell}
%=\frac{\ell^{2}}{\Rey_{p}U_{m}a}
%=\frac{\ell^{4}}{\Rey U_{m}a^{3}} \,.
%\end{equation*}
%%In our non-dimensionalized time $\tilde{t}$ 
%\ys{This} translates to a \ys{dimensionless} time $\tilde{t}\propto \kappa$.

Observe that our model of particle migration depends only on the cross-sectional coordinate $(\tilde{r}_{p},\tilde{z}_{p})$ and is independent of the current angular location within the curved duct $\theta_{p}$.
In order to study how far a particle travels through the curved duct over the time scale at which inertial migration takes place it is necessary to re-incorporate the axial motion into the model.
In practice, particles lag slightly from the surrounding fluid velocity.
However, for a small particle, this lag is sufficiently small that it is reasonable to take the background fluid velocity at the particle center as an approximation of the particle's axial velocity.
Therefore, we can incorporate this into our system of ordinary differential equations \eqref{eqn:migration_ode_alt} by adding
\begin{equation}\label{eqn:axial_ode_alt}%\begin{align}
\frac{d \tilde s_{p}}{d\tilde{t}} =\frac{d }{d\tilde{t}}(\mathrm{Re}_{c}\theta_{p}) = \frac{1}{(1+\ys{\epsilon\tilde{r}_{p}})}\frac{u_{1}\left(\tilde{r}_{p},\tilde{z}_{p}\right)}{U_{m}} \,,% \\
%\tilde s_p&=\mathrm{Re}_c\theta_p \label{eqn:axial_ode_alt_theta}%\\
%\frac{d\tilde s_{p}}{d\tilde{t}} &= \frac{u_{1}\left(\tilde{r}_{p},\tilde{z}_{p}\right)}{(1+\ys{\epsilon\tilde{r}_{p}})U_{m}} \,, 
%\end{align}
\end{equation}
where $\theta_{p}$ tracks the angular coordinate \ys{of the particle in} the curved duct and \ys{$\tilde s_p=\mathrm{Re}_c\theta_p$ is the corresponding dimensionless arc-length along the central axis of the channel that is related to the physical arc length} ${s}_{p}=R\theta_p$ \ys{by} %accumulates the distance travelled through the main axis 
%\ys{which %, for convenience, 
%has been} non-dimensionalized as 
\begin{equation}\label{eqn:axial dist, dim and nondim}
     \tilde s_{p}%=\frac{\ell}{2} \tilde s_{p}\Big/\left(\epsilon\Rey_{c}\right)
=\frac{\mathrm{Re}_{c}}{R} s_{p}\,.
%=\mathrm{Re}_{c} \theta_{p}\,.
\end{equation}
%\ys{[YS: the physical distance along the main axis of the duct will be $S=R\theta_p$ from which, using the same scaling for $S$ as for $s_p$ we find $d\tilde S/d\tilde t=\Rey_{c}d\theta_p/d\tilde t$. There seems little points in computing $\tilde S$ as it is just a scaling of $\theta_p$ and I am not sure if $\tilde s_p$ is really useful.]}
%[BH: perhaps $s_{p}$ would be better than $d_{p}$ here, i.e. since $d_{p}$ could be confused with the diameter of the particle, and using $s$ to measure length along a curve is reasonably common.]
\ys{We refer to $\tilde s_p$ as the distance the particle has travelled down the channel, where $\theta_p(0)=\tilde s_p(0)=0$. Finally, in keeping with the assumption made in developing this model, we take the $\epsilon\rightarrow 0$ limit of \eqref{eqn:axial_ode_alt}.}

To summarise, the complete ZeLF model is described by the first-order system of ordinary differential equations, \ys{involving just the single dimensionless parameter $\kappa$,}
\begingroup\allowdisplaybreaks %BH: I have allowed display breaks for this longish equation
\begin{subequations}\label{eqn:complete_model}\begin{align}
\label{eqn:ode of radial distance in a duct}\frac{d\tilde{r}_{p}}{d\tilde{t}} &= \frac{1}{6\pi\kappa}\tilde{r}_p\left(1-12.7\tilde{r}_p^{6}-24.8\tilde{z}_p^{6}\right)
\exp\left(2.95-1.43\tilde{r}_p^{2}-4.23\tilde{z}_p^{2}-1.98\tilde{r}_p^{4}\right. \,\\ &\hspace{2cm}\left.+5.28\tilde{r}_p^{2}\tilde{z}_p^{2}-1.10\tilde{z}_p^{4}+2.35\tilde{r}_p^{6}-1.16\tilde{r}_p^{4}\tilde{z}_p^{2}-7.16\tilde{r}_p^{2}\tilde{z}_p^{4}+3.51\tilde{z}_p^{6}\right) \notag\\
&\quad+0.01591(1-\tilde{r}_p^{2})^{2}(1-\tilde{z}_p^{2})\big(0.9833+0.2289\tilde{r}_p^{2} \notag\\
&\hspace{5cm}-4.9699\tilde{z}_p^{2}-1.1445\tilde{r}_p^{2}\tilde{z}_p^{2}+0.1246\tilde{z}_p^{4}\big) \,, \notag \\ 
%[KH: changeㅇ +1.1445 to -1.1445]
\frac{d\tilde{z}_{p}}{d\tilde{t}} &= \frac{1}{6\pi\kappa}\tilde{z}_p\left(1-24.8\tilde{r}_p^{6}-12.7\tilde{z}_p^{6}\right)
\exp\left(2.95-4.23\tilde{r}_p^{2}-1.43\tilde{z}_p^{2}-1.10\tilde{r}_p^{4}\right.\\ &\hspace{2cm}\left.+5.28\tilde{r}_p^{2}\tilde{z}_p^{2}-1.98\tilde{z}_p^{4}+3.51\tilde{r}_p^{6}-7.16\tilde{r}_p^{4}\tilde{z}_p^{2}-1.16\tilde{r}_p^{2}\tilde{z}_p^{4}+2.35\tilde{z}_p^{6}\right) \notag\\
&\quad+0.01591\tilde{r}_p(1-\tilde{r}_p^{2})\tilde{z}_p(1-\tilde{z}_p^{2})^{2}\big(3.4754+1.3734\tilde{r}_p^{2}-0.0712\tilde{z}_p^{2}\big)\,, \notag \\ 
%\frac{d \theta_{p}}{d\tilde{t}} &= \frac{1}{\ys{\mathrm{Re}_{c}(1+\epsilon\tilde{r}_{p})}}\left(1+0.1818(\tilde{r}_p^{2}+\tilde{z}_p^{2})\right)(1-\tilde{r}_p^{2})(1-\tilde{z}_p^{2}) \,, \\
\frac{d\tilde s_{p}}{d\tilde{t}} &= %\frac{1}{(1+\epsilon\tilde{r}_p)}
\left(1+0.1818(\tilde{r}_p^{2}+\tilde{z}_p^{2})\right)(1-\tilde{r}_p^{2})(1-\tilde{z}_p^{2})\label{eqn:axial_ode_alt_epsilon_limit} \,.
\end{align}\end{subequations}
\endgroup
%Here the constants $\kappa,\ys{\epsilon},\Rey_c$ are \ys{dimensionless} parameters of the model.
%\ys{[YS: removed a sentence here which didn't seem to me to be necessary; also added subscript $p$ to all $r$ and $z$ coords on RHS.]} %It can be observed that $\Rey_{c}$ simply applies a constant scaling factor to the angle travelled around the duct (similar applies to $d_{p}$ although this factor has been captured in the non-dimensionalisation $\breve{d}_{p}$).
%The influence of \ys{$\epsilon$} is negligible when it is \ys{small}, which was a general assumption in developing this model.
Results from \ys{this} model are straightforward to translate from dimensionless to dimensional coordinates as needed.

%We can perform another simple dimensional analysis to estimate how far a particle travels through a curved duct over the time scale $T_{m}$ over which migration occurs.
%In this case it is straight-forward to observe that the distance travelled down the main axis is $U_{m}T_{m}\propto \ell^{2}/(\Rey_{p}a)$ (or equivalently $\ell^{4}/(\Rey a^{3})=4R\kappa/\Rey$).
%In the context of our non-dimensionalized variable $\breve{d}_{p}$ this translates to a distance $\breve{d}_{p}\propto 2\kappa$.
%Consequently, the total angle travelled around the curved duct is anticipated to be proportional to $U_{m}T_{m}/(2\pi R)\propto \ell^{2}/(\Rey_{p}a R)$ (or equivalently $\ell^{4}/(\Rey a^{3}R)=4\kappa/\Rey$).
%\ys{[YS: dimensional analysis to give distance travelled along duct in migration time removed, given that dimensional analysis to give migration time was removed earlier.]}

\section{Results of the ZeLF model}\label{sec:toy_model_results}

\begin{figure}[t]
    \centering
    \begin{subfigure}[b]{0.975\textwidth}
        \centering
        \includegraphics[width=\textwidth]{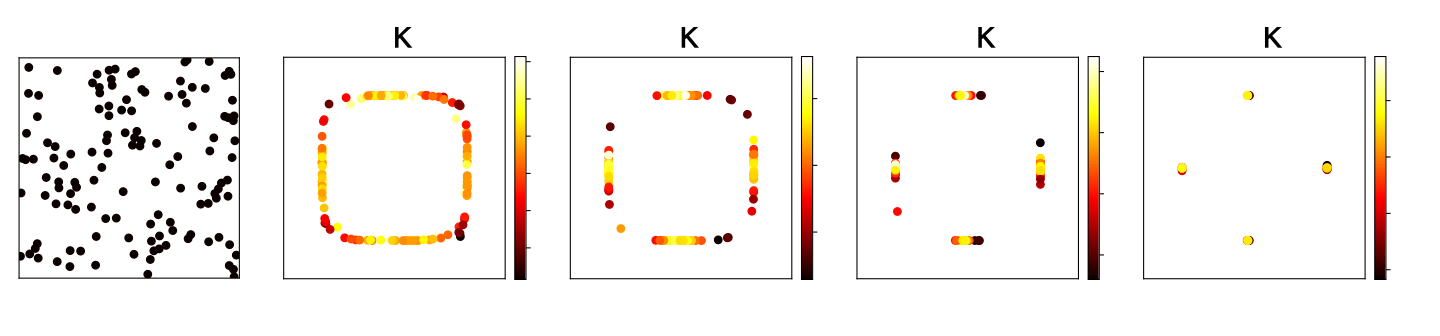}
        \caption{$\kappa= 1$}
        \label{fig: random initial points for kappa 1}
    \end{subfigure}
    \\
    \begin{subfigure}[b]{0.975\textwidth}
        \centering
        \includegraphics[width=\textwidth]{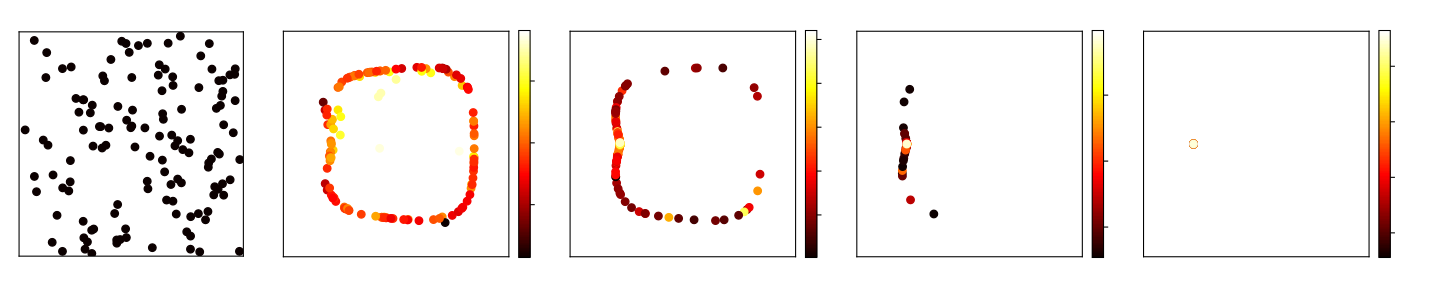}
        \caption{$\kappa= 25$}
        \label{fig: random initial points for kappa 25}
    \end{subfigure}
    \\ 
    \begin{subfigure}[b]{\textwidth}
        \centering
        \includegraphics[width=\textwidth]{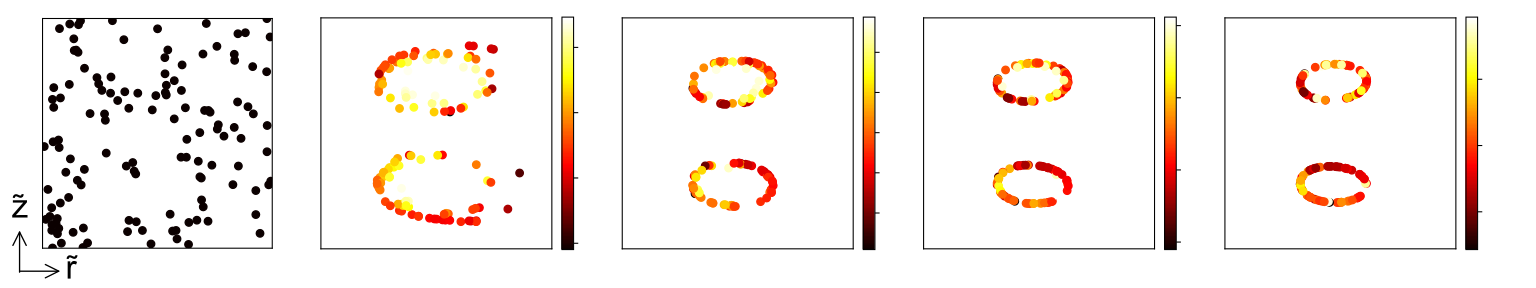}
        \caption{$\kappa= 200$}
        \label{fig: random initial points for kappa 200}
    \end{subfigure}
    \caption{Cross-sectional positions of 128 particles, initially randomly distributed within the cross-section, $[-1,1]\cross[-1,1]$. Each snapshot of the cross-section corresponds from left to right to the time $\tilde{t}=0,2\kappa,8\kappa,15\kappa,$ \red{and $30\kappa$}, and from top to bottom (a) small, (b) intermediate, and (c) large $\kappa$.
    The color scheme shows the axial distance each particle has traveled, calculated by \eqref{eqn:axial_ode_alt_epsilon_limit}.}
    \label{figs: random initial points for different kappa values}%PROJECT4
\end{figure}

In this section, we investigate particle motion and its dependence on $\kappa$ using the ZeLF model. 
Figure~\ref{figs: random initial points for different kappa values} illustrates the migration of particles via \red{five} snapshots in time for three distinct $\kappa$ values. The particles migrate towards a single fixed point ($\kappa=25$), one of multiple fixed points ($\kappa=1$) or to a stable orbit ($\kappa=200$).
The particle color indicates the \ys{dimensionless distance $\tilde s_p$ traveled down the channel by the particle.} %non-dimensionalized physical distanced travelled $\tilde{s}_p$ that each particle has completed, which satisfies
%\begin{equation}\label{eqn:axial_ode_alt_tilde_theta}
%    \frac{d \tilde{s}_{p}}{d\tilde{t}} \approx \frac{u_{1}\left(\tilde{r}_{p},\tilde{z}_{p}\right)}{U_{m}}.
%\end{equation}
%This is a simplification of  \eqref{eqn:axial_ode_alt_theta} under the assumption that the duct size $l$ is small relative to the duct radius $R$ \ys{($\epsilon\rightarrow 0$)}.

We observe three distinct behaviors, \ys{here termed ``multi-focus'', ``unique-focus'' and ``periodic orbit'', corresponding to $\kappa$ small ($\kappa\lesssim 10$), intermediate ($10\lesssim\kappa\lesssim 25$), and large ($\kappa\gtrsim 25$).} %, which we denote \ys{by the} ``Multi-focus region", ``Unique focus region", and ``Periodic orbit region" respectively (the numbers associated with the $\kappa$ values are not the definite[KH: not sure how to make it more clear. I am trying to indicate that the numbers 10, 25 are not pinpoint accurate but indicating that the behavior changes with in the range 9~11 and 24~26] there values).
% \ys{[I've used $\lesssim,\gtrsim$ and deleted the comment, which I think works.]}
Since we are interested in the long term behavior of the particles we only \ys{discuss} %talk about 
the long-time limit sets ($\omega$-limit sets \cite{guckenheimer2013nonlinear}). 
The limit sets are composed of equilibria, which are classified as stable nodes \ys{or foci}, saddle points, and unstable nodes by their eigenvalues, and limit cycles, also classified as either stable or unstable by their \Poincare\ map.

\begin{figure}[t]
    \includegraphics[width=\textwidth]{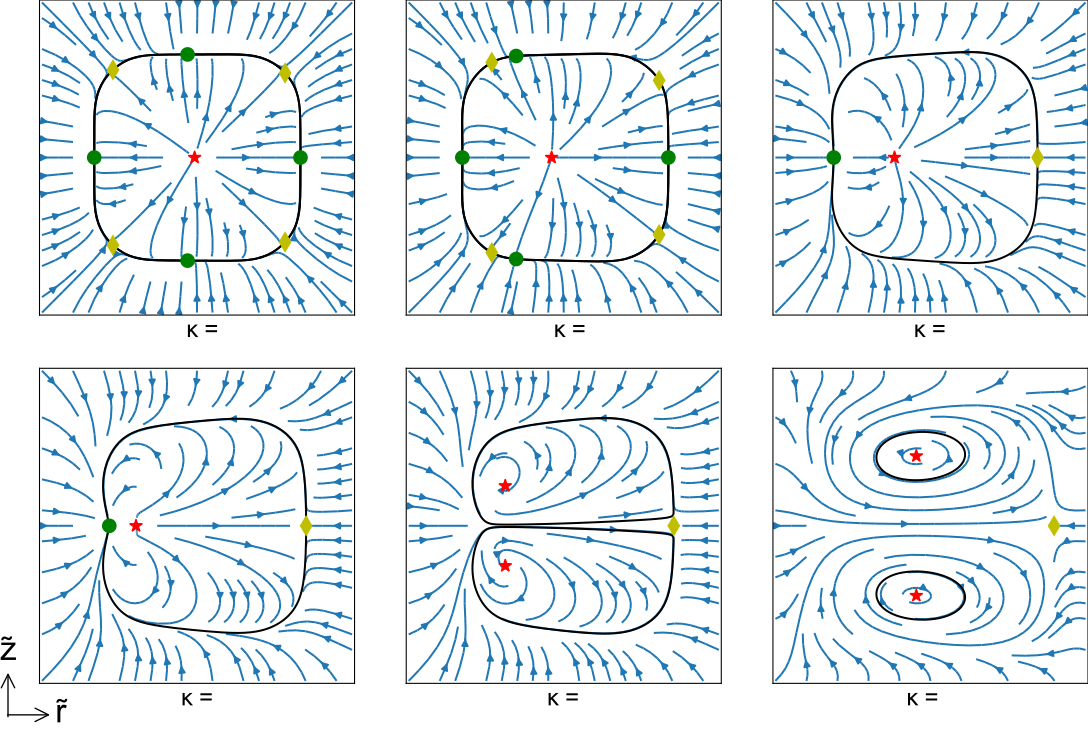}
    \caption{Particle trajectories for different $\kappa$ values within the cross-section, $[-1,1]\cross[-1,1]$. The equilibria are in different colors and shapes: {\color{green!30!black}stable nodes (green $\bullet$)}, {\color{yellow!50!black} saddle points (yellow $\diamondsuit$)}  and {\color{red!90!black}unstable nodes (red $\star$)}.
    For $\kappa\leq25$, the black line represents the heteroclinic orbit that connects the saddle to the stable equilibria. For $\kappa\geq30$, the black line represents the limit cycle.}
    \label{figs: postition of eq pts and limit cycles for large kappa values}
\end{figure}

\subsection{Multi-focus \ys{behavior} %region 
(small $\kappa$)}

\begin{figure}[t]
    \centering
    \begin{subfigure}[b]{0.45\textwidth}
        \centering
        \includegraphics[width=\textwidth]{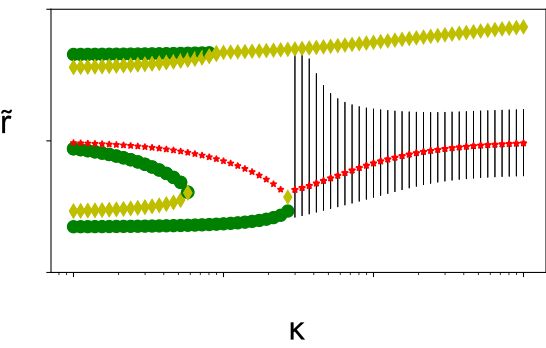} \caption{$\kappa$-$\tilde{r}$ position}
        \label{fig: kappa-r position stable and limit cycle only}
    \end{subfigure}
    \hfill
    \begin{subfigure}[b]{0.45\textwidth}
        \centering
        \includegraphics[width=\textwidth]{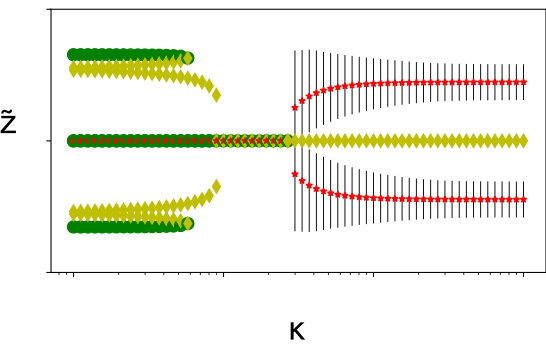}
        \caption{$\kappa$-$\tilde{z}$ position}
        \label{fig: kappa-z position stable and limit cycle only}
    \end{subfigure}
    \caption{Equilibrium positions and the limit cycle range as a function of $\kappa$. The equilibria are in different colors and shapes: {\color{green!30!black}stable nodes (green $\bullet$)}, {\color{yellow!50!black} saddle points (yellow $\diamondsuit$)}  and {\color{red!90!black}unstable nodes (red $\star$)}. The range of the limit cycles are shown in black vertical lines.}
    \label{figs: kappa and position graphs stable and limit cycle only}
\end{figure}

For small $\kappa$, there are multiple \ys{stable nodes or}
focusing points near the center of each side of the cross-section, multiple saddle points near the corners of the cross-section, and one unstable node in the middle (Figure~\ref{figs: postition of eq pts and limit cycles for large kappa values}{\color{green!50!black}ab}). 
For each saddle point, there is a heteroclinic orbit that connects to a stable node which acts as a slow manifold. 
The particles quickly migrate onto one of these heteroclinic orbits, and then slowly converge to the stable node (Figure~\ref{fig: random initial points for kappa 1}).
Therefore, the migration velocity of the particle on the slow manifold determines the time needed for the particles to converge to the stable nodes.
These particle trajectories are similar to those in a straight duct which is expected since as $R\rightarrow \infty$,  $\kappa\rightarrow 0$ \cite{Harding2019JFM,Hood2016RSC}. 

\subsection{Unique focus \ys{behavior} %region
(intermediate $\kappa$)}

\begin{figure}[t]
    \centering
    \begin{subfigure}[b]{0.49\linewidth}
        \centering
        \includegraphics[height=0.6\textwidth]{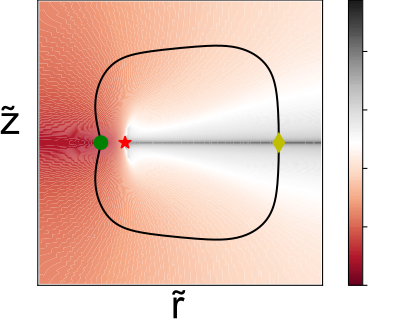}
        \caption{}
        \label{fig: heat map of distance to converge, kappa = 25}
    \end{subfigure}
    \hfill
    \begin{subfigure}[b]{0.49\linewidth}
        \includegraphics[height=0.6\textwidth]{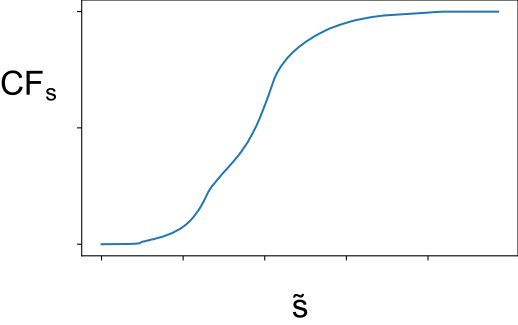}
        \caption{}
        \label{fig: cumulative hist of distance to converge, kappa = 25}
    \end{subfigure}
    \\
    \begin{subfigure}[b]{0.49\linewidth}
        \centering
        \includegraphics[height=0.6\textwidth]{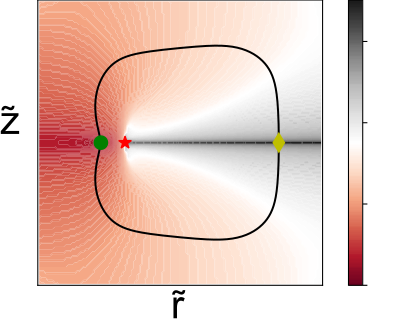}
        \caption{}
        \label{fig: heat map of time to converge, kappa = 25}
    \end{subfigure}
    \hfill
    \begin{subfigure}[b]{0.49\linewidth}
        \includegraphics[height=0.6\textwidth]{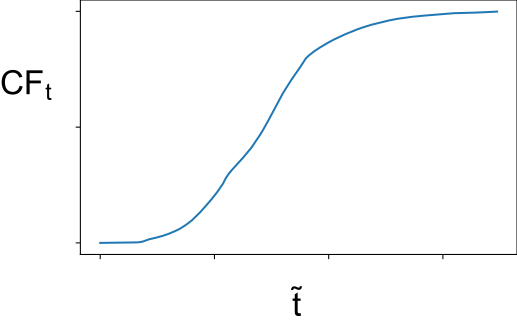}
        \caption{}
        \label{fig: cumulative hist of time to converge, kappa = 25}
    \end{subfigure}
    \caption{On the left are heat maps of the cross-section, $[-1,1]\cross[-1,1]$, that show the (a) axial distance $\tilde s_p^*$ and (c) time $\tilde{t}^*$, \ys{defined} in \eqref{eq: definition of axial distance and time required to focus}, required to focus from a position in the cross-section to the stable equilibrium point for $\kappa= 25$. The equilibria are shown in different colors and shapes: {\color{green!30!black}stable nodes (green $\bullet$)}, {\color{yellow!50!black} saddle points (yellow $\diamondsuit$)}  and {\color{red!90!black}unstable nodes (red $\star$)}.    The black solid line represents the heteroclinic orbit connecting the saddle to the stable node. On the right, the graphs show \ys{(b) $CF_s$ versus $\tilde s$ and (d) $CF_t$ versus $\tilde t$ as defined in \eqref{eq: definition of cumulative axial distance and time required to focus}, i.e.} the fraction of the cross-sectional area \ys{from} which particles \ys{have focused} within the given distance $\tilde s$ and time $\tilde t$. }%project 3
    \label{figs: Heat map indicating the distance required to converge for different kappa values}
\end{figure}

\begin{figure}[t]
    \centering
    \begin{subfigure}[b]{0.47\textwidth}
        \centering
        \includegraphics[width=\textwidth]{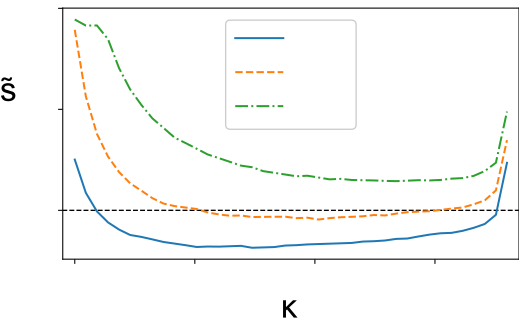}
        \caption{}\label{fig: multiple kappa and distance required}
    \end{subfigure}
    \hfill
    \begin{subfigure}[b]{0.47\textwidth}
        \includegraphics[width=\textwidth]{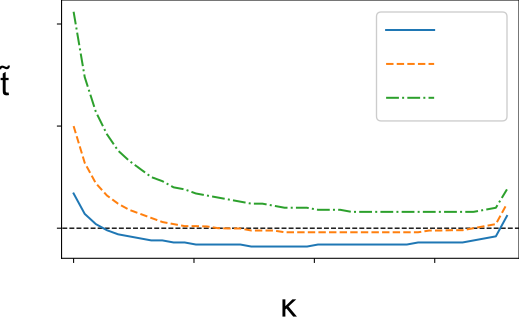}
        \caption{}\label{fig: multiple kappa and time required}
    \end{subfigure}
    \caption{(a) axial distance $\tilde{s}$ and (b) time $\tilde{t}$ required for 
    {\color{blue!80!black}90\% (blue)},
    {\color{red!80!black}95\% (red)},
    {\color{green!30!black}99\% (green)} of the particles to focus to the equilibrium for intermediate $\kappa$. \red{The dashed horizontal line indicates the distance and time at which approximately 95\% of particles with $15\leq\kappa\leq25$ are focused.}}
    \label{figs: multiple kappa and distance or time required}
\end{figure}

As $\kappa$ increases \ys{the system undergoes saddle-node bifurcation} as \ys{the stable node and saddle point in each of the upper and lower halves of the duct} %two pairs of stable nodes and saddle points (located on the top and bottom of the duct) 
merge and disappear. \ys{Simultaneously a (subcritical) pitchfork bifurcation happens as} %while 
two saddle points and one stable node on the outer side (right side) merge and become a saddle point (Figure %\ref{figs: postition of eq pts and limit cycles for large kappa values}b,\ref{figs: postition of eq pts and limit cycles for large kappa values}c, and 
\ref{figs: kappa and position graphs stable and limit cycle only}). \ys{This results} in a system with a unique stable equilibrium point that attracts all particles in the duct  (Figure~\ref{figs: postition of eq pts and limit cycles for large kappa values}{\color{green!50!black}cd}). 
All equilibria lie on the $\tilde{r}$-axis, due to the \ys{vertical} reflection symmetry, with the stable node on the inner side (left side) of the duct.
Similar to the multi-focus behavior, %in the Multi-focus region, 
there exist heteroclinic orbits that connect the saddle to the stable node which act as a slow manifold (Figure~\ref{fig: random initial points for kappa 25}). %\ref{figs: postition of eq pts and limit cycles for large kappa values}d).
% In addition to the saddle and stable node, there is an unstable node, which lies on the $\tilde{r}$-axis due to the reflection symmetry . 
% The $\tilde{r}$-coordinate of the stable node is in the range $[-0.75,-0.5]$, and gets closer to the center of the duct as $\kappa$ increases (Figure~\ref{figs: kappa and position graphs stable and limit cycle only}). 

The axial distance and time required for particles to focus on this equilibrium point are important since \ys{they determine} the length and run time of the apparatus to achieve particle focusing.
Technically, if a particle does not start on the stable equilibrium point $(\tilde{r}^*,\tilde{z}^*)$ it will take infinite time to arrive at the exact equilibrium point. 
\ys{However,} we consider that a particle has ``focused" at the equilibrium point if the distance from the particle to the equilibrium is smaller than a certain threshold, which we choose \ys{to be} $0.01$ for this paper. 
We define 
\begin{subequations}\label{eq: definition of axial distance and time required to focus}\begin{align}
    \tilde{t}^*(\tilde{r},\tilde{z}) &= 
    \min\big\{\tilde{t}:
    \left\lVert(\tilde{r}_{p},\tilde{z}_{p})(\tilde{t})
    -(\tilde{r}^*,\tilde{z}^*)\right\rVert<0.01, \,
%    (\tilde{r}_{p},\tilde{z}_{p},\tilde{s}_{p})(0)=
%    (\tilde{r},\tilde{z},0)
    (\tilde{r}_{p},\tilde{z}_{p})(0)=
    (\tilde{r},\tilde{z})
    \big\},\\
    \tilde{s}_p^*(\tilde{r},\tilde{z}) &= \tilde{s}_{p}(\tilde{t}^*(\tilde{r},\tilde{z})),%\kh{\text{with initial point}} \ys{\, \tilde s_p(\tilde r,\tilde z)=0,}
    % \breve{d_p}^*(\tilde{r},\tilde{z}) &= 
    % \min\big\{\breve{d}_{p}(\tilde{t}):
    % \left\lVert(\tilde{r}_{p},\tilde{z}_{p})(\tilde{t})
    % -(\tilde{r}^*,\tilde{z}^*)\right\rVert<0.01, \,
    % (\tilde{r}_{p},\tilde{z}_{p},\breve{d}_{p})(0)=
    % (\tilde{r},\tilde{z},0)
    % \big\}
\end{align}
\end{subequations}
which correspond to the required \ys{time and} axial distance for a particle at an initial position  $(\tilde{r},\tilde{z})$ in the cross-section to focus. 
The heat maps in Figure~\ref{figs: Heat map indicating the distance required to converge for different kappa values}{\color{green!50!black}ac} show, $\tilde{s}_p^*$ and $\tilde{t}^*$ \ys{over} %for different initial position in 
the cross-section for $\kappa = 25$. 
%Particles that start in the red regions require the shorter axial distance and time to focus while the black regions require longer axial distance and time to focus.
\ys{In each map, black shows the region from which the focusing distance or time is greatest, and dark red shows the region from which the focusing distance or time is shortest.}

In order to understand the overall focusing ability \ys{in a given axial length or time, we compute the fraction of the channel cross-section area from which particles will have focused in distance $\tilde s$ or time $\tilde t$.} We define these functions $CF_{s}(\tilde{s})$ and $CF_{t}(\tilde{t})$, respectively, \ys{as:}
\begin{subequations}
\label{eq: definition of cumulative axial distance and time required to focus}
\begin{align}
    CF_s(\tilde{s}) &=
    \frac{1}{4}\int_{-1}^{1}\int_{-1}^{1} \scalebox{1.5}{$\chi$}_{\tilde{s}_p^*\leq\tilde{s}}(\tilde{r},\tilde{z})d\tilde{r}d\tilde{z},\\
    CF_t(\tilde{t}) &=
    \frac{1}{4}\int_{-1}^{1}\int_{-1}^{1} \scalebox{1.5}{$\chi$}_{\tilde{t}^*\leq\tilde{t}}(\tilde{r},\tilde{z})d\tilde{r}d\tilde{z},
\end{align}
\end{subequations}
where \red{$\scalebox{1.5}{$\chi$}_{\tilde{s}_{p}^*\leq\tilde{s}}$ and $\scalebox{1.5}{$\chi$}_{\tilde{t}^*\leq\tilde{t}}$} are characteristic functions with \ys{unit} value \ys{where} $\tilde{s}_{p}^*(\tilde{r},\tilde{z})\leq\tilde{s}$ and $\tilde{t}^*(\tilde{r},\tilde{z})\leq\tilde{t}$, respectively, \ys{and which are zero elsewhere}.
Assuming that the particles are initially randomly \ys{distributed} in the cross-section, \ys{$CF_s$ and $CF_t$ give good approximations} of the fraction of particles focused after a given distance $\tilde{s}$ and time $\tilde{t}$.
Figure~\ref{figs: Heat map indicating the distance required to converge for different kappa values}{\color{green!50!black}bd} plot $CF_{s}(\tilde{s})$ and $CF_{t}(\tilde{t})$ for $\kappa=25$.
We observe that the plots of $CF_{s}$ and $CF_{t}$ are qualitatively similar.
This is due to the fact that the change of axial velocity is relatively slow and smooth over the slow manifold.

% [KH: I think this paragraph needs help]
For engineering design one
needs to know the duct length/number of turns of the spiral 
 the \ys{duct length} %distance and time 
required for the majority \ys{(e.g. $95\%$)} of particles to focus. This corresponds to finding $\tilde{s}$ \ys{satisfying} %and $\tilde{t}$ that satisfy 
$CF_{s}(\tilde{s})=f$, % and $CF_{t}(\tilde{t})=f$ respectively, 
\ys{where $f$ is the fraction of particles focused.}
Figure~\ref{figs: multiple kappa and distance or time required} %{fig: multiple kappa and distance required}
shows for $10\leq\kappa\leq28$ the axial distance \ys{and time} required for $90$, $95$, and $99\%$ of the particles to focus. \ys{There is only weak dependence on $\kappa$ for a significant portion of this range. These plots show that a given duct geometry may be used to focus a range of particle sizes. For example, from Figure~\ref{fig: multiple kappa and distance required}, we see that a duct of length $\tilde s=300$ will focus at least 95\% of particles corresponding to $15\leq\kappa\leq 25$.}
Once we know the duct length we can estimate how many turns of the spiral required for the focusing. Note that the bend radius is not constant along the spiral however the duct width is much smaller than the bend radius, allowing for small variation in the bend radius in real devices (see e.g. Fig.~\ref{fig:physical photo of spiral duct}). 
As an example, using physical parameters in \cite{ramachandraiah2017inertial,warkiani2016ultra}
 the number of rotations corresponding to $\tilde s=300$ is approximately 4 and 1-5, \red{respectively. We note that the duct cross-sections used in the experiments are} rectangular rather than square.  Still this suggests that only a modest number of turns may be needed to achieve the desired focusing.
\ys{Then, since the particle radius is $a= G/\kappa^{1/3}$, where $G=(\ell^4/4R)^{1/3}$ captures the other geometrical parameters of the duct, at least 95\% of all particles of size \[0.34G\approx G/25^{1/3}\leq a\leq G/15^{1/3}\approx 0.41G\] will focus after passing through the duct.
From Figure~\ref{fig: multiple kappa and time required} we see that this focusing takes a time $\tilde t\approx 500$. Differences in the size of particles}
%The axial distance is weakly dependent on $\kappa$ between the values of $15$ and $25$. 
%This indicates that for the same physical setup there is a range of particle sizes that focus within similar distances. 
%For example, if we set up the experiments so that the particle with $\kappa = 20$ focus, then particles with $\kappa\in[15,25]$ will also focus within a similar distance, resulting in particles with radius $\pm10\%$ of the target particle accumulated.
%This indicates that the same duct can be used to focus similar-sized particles at the end of the duct. 
%The difference in the size of the particles 
results in different focusing points %which can be used to our advantage to sort the particles.
\ys{and, hence, particle separation by size.  In practice, there needs to be a significant separation, within the cross section, between focusing positions (as a function of particle size) for effective separation. For example, in Figure~\ref{fig: kappa-r position stable and limit cycle only}, the variation in the radial coordinate is not large so that separation may be difficult. In general, for square ducts, the focusing position may not have sufficient variation and other cross-sectional shapes may be  preferred as discussed in \cite{Harding2019JFM}}.
An analysis similar to this work could be carried out for different cross-sections.
%\kh{However, the difference of the position of the focusing point in this $\kappa$ range is not dramatic, as shown in Figure~\ref{fig: kappa-r position stable and limit cycle only}, making the separation of particles rather tricky.}
% \ys{[Discuss sensitivity of focusing point in this $\kappa$ range and refer to figure 8(a). In principle can separate particles but, for square duct the focusing point is not very sensitive to $\kappa$ so separating particles might be tricky.]}
%Repeating the calculations for time (Figure~\ref{fig: multiple kappa and time required}), gives us qualitatively similar results.

\subsection{Periodic orbit \ys{behavior} %region 
(large $\kappa$)}

\begin{figure}[t]
    \centering
    \begin{subfigure}[h]{0.55\textwidth}
        \centering
        \includegraphics[height=7cm]{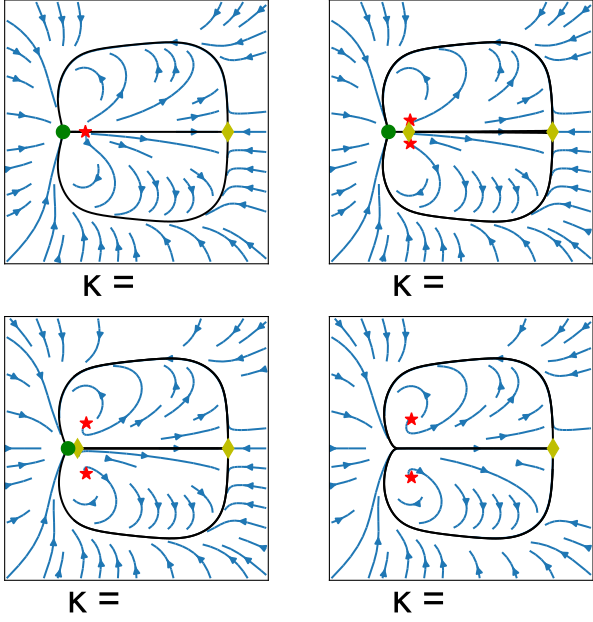}
        \caption{Particle trajectories and equilibria}
        \label{fig: Fluid flow between 25 30}
    \end{subfigure}
    \hfill
    \begin{subfigure}[h]{0.44\textwidth}
        \centering
        \includegraphics[height=7cm]{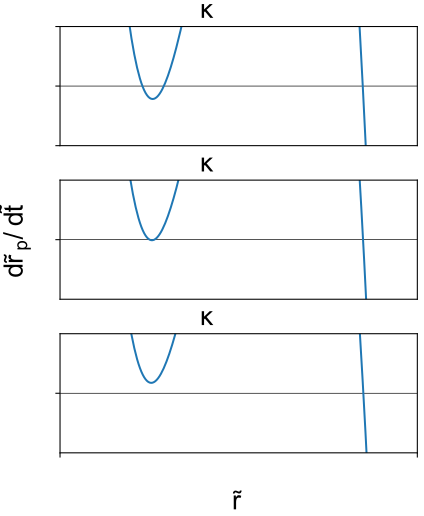}
        \caption{${d\tilde{r_p}}/{d\tilde{t}}$ on the $\tilde{r}$ axis}
        \label{fig: Fluid flow on the hline between 25 30}
    \end{subfigure}
    \caption{The two bifurcations between $\kappa=25$ and $30$ within the cross-section, $[-1,1]\cross[-1,1]$. (a) presents the particle trajectories and the equilibria in different colors and shapes: {\color{green!30!black}stable nodes (green $\bullet$)}, {\color{yellow!50!black} saddle points (yellow $\diamondsuit$)}  and {\color{red!90!black}unstable nodes (red $\star$)}.
    For $\kappa\leq28$, the black line represents the heteroclinic orbit that connects the equilibria on the $\tilde{r}$ axis. For $\kappa=28.5$, the black line represents the limit cycle. 
    (b) are graphs of ${d\tilde{r_p}}/{d\tilde{t}}$ on the $\tilde{r}$ axis for $28\leq\kappa\leq28.2$}
    % $[-5,5]\cross10^{-4}$ are shown since we are interested in the bifurcation.}
    \label{figs: fluid flow between 25 30}
\end{figure}

As $\kappa$ increases \ys{further, the system undergoes two bifurcations as the stable and unstable nodes on the $\tilde r$-axis collide and give birth to an unstable node on either side of the $\tilde r$-axis (Figure~\ref{fig: Fluid flow between 25 30}).} 
For $\kappa$ between 25 and 25.5, there exists a vertical pitchfork bifurcation where the unstable node bifurcates to a pair of unstable nodes and a saddle point between them. For $\kappa$ between 27.5 and 28.5, there exists a horizontal saddle-node bifurcation where the saddle point and the stable node cancel out each other.
Since ${d\tilde{z_p}}/{d\tilde{t}}=0$ on the $\tilde{r}$-axis, the latter bifurcation is determined by ${d\tilde{r_p}}/{d\tilde{t}}$ on the axis, which is shown in Figure~\ref{fig: Fluid flow on the hline between 25 30}.
After the two bifurcations occur, there remains two periodic periodic orbits on either side of the $\tilde{r}$-axis, \ys{each around one of the unstable nodes} (Figure~\ref{figs: postition of eq pts and limit cycles for large kappa values}{\color{green!50!black}e}).
% the heteroclinic $\tilde{r}$ axis
% After the two bifurcations occur, the heteroclinic orbit that connects the saddle and the stable node, and two that connect the equlibria on the $\tilde{r}$ axis, merge \ys{to} produce
%The stable and unstable nodes merge and disappear while two unstable nodes appear within the two periodic orbits.
For $\kappa=30$, the periodic orbits are large with varying particle speeds, slower near the saddle point and where the saddle-node bifurcation \red{occurs}. 
As $\kappa$ increases the periodic orbits \ys{become} smaller and the particle speed becomes more uniform as they effectively follow the Dean flow (Figure~\ref{figs: postition of eq pts and limit cycles for large kappa values}{\color{green!50!black}f}, and \ref{figs: kappa and position graphs stable and limit cycle only}).

% \kh{
% We also take a deeper look at the bifurcations between $\kappa$, 25 and 30 (Figure~\ref{figs: postition of eq pts and limit cycles for large kappa values}{\color{green!50!black}de}). Figure~\ref{fig: Fluid flow between 25 30} shows that between the two $\kappa$ values there are two bifurcations happening consecutively. } 

\begin{figure}[t]
    \centering
    \begin{subfigure}[h]{0.45\textwidth}
        \centering
        \includegraphics[width=\textwidth]{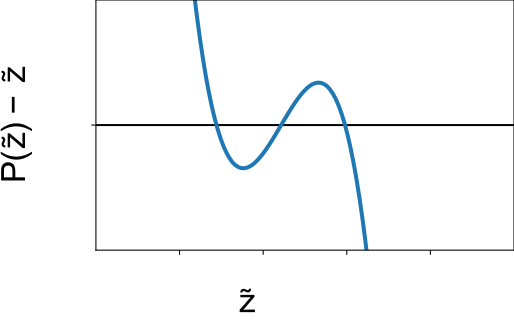}
        \caption{Poincare map for $\kappa$ = 200}
        \label{fig: poincare map}
    \end{subfigure}
    \hfill
    \begin{subfigure}[h]{0.45\textwidth}
        \centering
        \includegraphics[width=\textwidth]{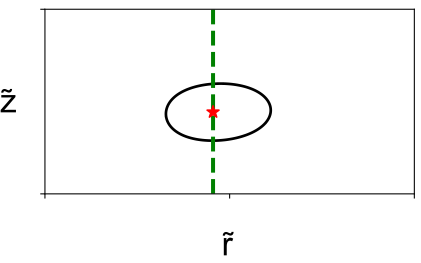}
        \caption{Limit cycle and the manifold}
        \label{fig: poincare map explain}
    \end{subfigure}
    \caption{The \Poincare{} map \ys{$P(\tilde z)$} of the upper half domain, $[-1,1]\cross[0,1]$, for $\kappa = 200$. The manifold we choose for the \Poincare{} map is the vertical line that passes through the {\color{red!90!black}unstable nodes (red $\star$)} as shown in (b) and (a) is the difference \ys{between $\tilde z$ and its \Poincare{} map $P(\tilde z)$,} %of these two z-coordinates after we apply the \Poincare{} map,
    i.e. $P(\tilde{z})-\tilde{z}$. }
    \label{figs: poincare map}
\end{figure}

% [BH: I think we could probably elaborate a bit more in the proof and possibly even prove something a little stronger. 
% For example: proving there is $\kappa$ large enough that $v_{r}(r,0)>0$ on $r\in(-1,0)$:
% $D_{r}(r,0)>0$ on $r\in(-1,0)$, $L_{r}(r,0)>0$ on an interval $(-1,r_{0})$ and $L_{r}(r,0)<0$ on an interval $(r_{0},0)$ ($r_{0}<0$ such that $L_{r}(r_{0},0)=0$). We can define $\alpha=\min\{D_{r}(r,0):r\in(r_{0},0)\}$ and  $\beta=-\min\{L_{r}(r,0):r\in(-1,0)\}$, both $\alpha,\beta$ are positive and finite, then for $\kappa>\beta/\alpha$ we have $v_{r}(r,0)>0$ for all $r\in(-1,0)$ (we could even estimate $\alpha,\beta$).
% Can we then go on to continue this type of reasoning to prove the existence of the different equilibrium rather assume so in the statement of the theorem?]

% [: I think the existence of the saddle point might be possible. I am worried about the unstable nodes though. If we do not assume the existence of them Poincare-Bendixson is not that useful, as we cannot rule out that there are stable equilibria on the upper and lower halves of the cross-section. So I think we need to assume the existence of unstable nodes (or at least that there aren't any stable nodes). In this view I don't think it will add much to prove that there exists a saddle. If we have to check if there aren't any stable nodes that is almost identical  as checking if there are saddle as we have to check if there are equilibria and what is their stability.] 

These periodic orbits are attractive limit cycles that are unique on each half domain, as shown by calculating the \Poincare{} map \cite{guckenheimer2013nonlinear}. 
We choose the manifold that defines the \Poincare{} map as the vertical line through the unstable equilibrium point \ys{in the top half of the duct} (Figure~\ref{fig: poincare map explain}) and define the \Poincare{} map $P:(0,1)\rightarrow(0,1)$ \ys{taking as input} %such that it inputs 
the $\tilde{z}$-coordinate %value 
of \ys{a particle} %the point 
on the manifold and \ys{yielding as} output the $\tilde{z}$-coordinate %value once it makes 
\ys{of that particle after} a full rotation around the unstable equilibrium. 
Figure~\ref{fig: poincare map} shows the $P(\tilde{z})-\tilde{z}$ value for the variable $\tilde{z}$.
There are three zeros of the function, the middle one corresponding to the unstable node of the system and the other two corresponding to the limit cycle. 
The sign on either side of the zeros indicates that this limit cycle is attractive.

\subsection{\ys{Comparison of ZeLF and detailed numerical models}}\label{sec:dynamics_comparison}

\begin{figure}[t]
%\psfrag{kappa}[t][t]{$\kappa$} % <--- YS: use of psfrag to substitute figure text
    \centering
    \begin{subfigure}[t]{0.9\textwidth}
        \mbox{}\\
        \centering
        \includegraphics[height=6cm,keepaspectratio]{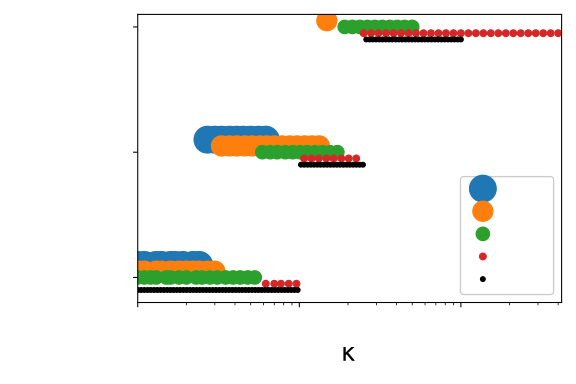}
    \end{subfigure}
    % \caption{$\kappa$ values and particle flow Behavior of the Actual Model. The flow behavior is classified according to the characteristics given in section \ref{sec:dynamics_comparison}. For three points the positions of the equilibria are shown in the boxes that the arrows connect. The green, yellow and red circles correspond to the {\color{green!30!black}stable nodes},
    % {\color{yellow!50!black}saddles}
    % and {\color{red!90!black}unstable nodes} respectively.}
    \caption{$\kappa$ values and particle flow behavior of the numerical Model. The flow behavior is classified as described in section \ref{sec:dynamics_comparison}. The different sized circles indicate the four different sized particles from $a=0.05$ to $0.2$. The black line indicates the flow behavior of the ZeLF model. As the size of the particles decrease the behavior matches that of the ZeLF model. Due to the restriction on the range of $R$, the data points of particles with size $a=0.15$ and $0.2$ do not fully extend to exhibit the periodic flow behavior. 
    }%PROJECT 7
    \label{fig: kappa and Behavior of Brendans data}
\end{figure}

\ys{Recall that the ZeLF model used for the above analysis is an approximation of the detailed numerical model of \cite{Harding2019JFM}. We here compare the predictions of particle dynamics of these two models. From the ZeLF model, we have found that the particle dynamics will differ depending on the value of the parameter $\kappa$. For small $\kappa$ $(\lesssim 10)$ there are multiple points in the duct cross-section to which particles focus depending on their initial position; for intermediate $\kappa$ $(10\lesssim\kappa\lesssim 25$) there is a unique point to which all particles will focus, regardless of their initial position; for large $\kappa$ $(\gtrsim 25)$ there are no stable focus points but particles initially located in the top/bottom half of the duct cross-section will migrate to a periodic orbit in the top/bottom half of the cross-section. Similar changes in particle dynamics are seen using the detailed numerical model but these occur at slightly different values of $\kappa$.

Using a square duct with parameters $l=2$, $80\leq R\leq 5120$ and $a=0.05,0.10,0.15,0.20$, Figure~\ref{fig: kappa and Behavior of Brendans data} shows the nature of the particle dynamics predicted by the detailed numerical model of \cite{Harding2019JFM} for different values of $\kappa$. Also shown, for comparison, are the predictions of the ZeLF model which assumes that the particle size is small compared to the size of the duct ($a\ll \ell$) and that the bend radius $R$ is large ($\epsilon\rightarrow 0$). There is excellent agreement between the ZeLF model and the numerical model for small particle size. As the particle size increases the numerical model gives transitions between the different behaviors at smaller values of $\kappa$, but qualitatively we continue to observe the same three regimes occurring in the same order.}

\section{Conclusion}\label{sec:Conclusions}
%This paper generates a model for studying the particle behavior that is flowing down a curved duct. 
%Building on previous work \cite{Harding2019JFM} the model we develop depends only on $\kappa$ which is a parameter that combines different length parameters to explain the particle flow behavior. 
%[K: Brendan can you write Some comments about building the model] 
%Once we build the model we use it to understand the dynamics of the particle flow.
%%% BH: Kyung, I have written the following paragraph, not sure if this is what you had in mind.
Building on previous work~\cite{Harding2019JFM} we \ys{have developed} \red{a} simplified model for the migration of a small neutrally buoyant particle suspended in relatively slow flow through \ys{microfluidic} curved ducts \red{with} a square cross-section.
While curved ducts with a square cross-section %, compared to a rectangular cross-section, 
are not as \ys{effective as rectangular (or other) cross-sections for particle separation by size,} 
%useful in many of the applications of inertial microfluidics
they exhibit a wide range of interesting  \kh{bifurcations}.
%We approximate the inertial lift force by first fitting the zero level curves for a straight duct and before fitting the magnitude elsewhere. \ys{(This needs more detailed description.)} This is essential to ensure the correct migration behavior for a very large bend radius.
\bh{We model the inertial lift force by first fitting the zero level curves to data obtained from simulations of a small particle in a straight duct. Then, multiplying by the exponential of a polynomial in the cross-section coordinates, we fit the inertial lift force over the entire cross-section in a manner that does not modify the zero level curves already modeled. This two-step process is essential to capture the correct topology of the inertial lift force and correctly predict equilibria for large bend radii.}
To this, we add a simple drag force model to capture the effect of the secondary motion of the background flow.
The ratio of the secondary drag to inertial lift forces is parametrized by a single dimensionless variable $\kappa$.
Unlike previous studies, we also incorporate \ys{travel along the duct} %the behavior down the main axis 
into the trajectory model \ys{to enable an analysis of} %so that we may analyze 
the time and distance required for particles to focus.

Using this model we observe that there exist three different $\kappa$ regions with distinctive flow behavior. 
We introduce a simple criterion to categorize these three $\kappa$ behavior. 
This categorization aids in identifying appropriate ranges of physical parameters when designing a curved duct for focusing purposes. 
The methodology and analysis applied to extract and understand the $\kappa$ dependence of the model can be applied to other shaped ducts as well. 
In addition, \ys{we have shown that a duct of a given length will focus particles over} a range of $\kappa$ values. %, the axial distance over which focusing occurs is relatively similar. 
This is an important observation as it establishes that a single device design \ys{might} be used to focus multiple particle sizes simultaneously. 
Going forwards, it will be interesting to study if this observation holds for non-square ducts and/or non-spherical particles \ys{such that the focusing points for particles of different sizes have sufficiently different radial coordinate to enable practical particle separation}.

%\appendix
%\section{ERASE IF NOT USED} 

\renewcommand{\thefootnote}{\fnsymbol{footnote}}
\section*{Acknowledgments}
\red{We thank the Warkiani Laboratory at the University of Technology Sydney (https://www.warkianilab.com) for providing the photo in Figure 1(b), and Prof. Dino Di Carlo at the University of California, Los Angeles for useful suggestions.}
% Funding is at the start of the paper.

% \section*{Potential referees} 
% Kaitlyn Hood,  Jeff Morris, Sungyon Lee,
% Jonathan Wiley (probably not, Yvonne has published within 4 years), Howard Stone (Probably busy), John Wang, 
% Huaxiong Huang (York), Alejandro Martínez-Calvo (Postdoc of Howard Stone)

% Editorial board : Jeff Moehlis (1st), Theodore Kolokolnikov, Martin Wechselberger, James Sneyd (? probably not), Gary Froyland, Marty Golubitsky

\bibliographystyle{siamplain}
\bibliography{mybib}
\end{document}